\documentclass{emulateapj}

\slugcomment{ApJ, 2009, in press}

\begin{document}

\shortauthors{Gordon et al.}
\shorttitle{FUSE Extinction Curves}

\title{{\it FUSE} Measurements of Far-Ultraviolet Extinction. III.
The Dependence on R(V) \\ and
Discrete Feature Limits from 75 Galactic Sightlines}

\author{Karl~D.~Gordon}
\affil{Space Telescope Science Institute, 3700 San Martin
Drive, Baltimore, MD 21218, kgordon@stsci.edu}
\and
\author{Stefan~Cartledge, Geoffrey~C.~Clayton}
\affil{Dept.\ of Physics \& Astronomy, Louisiana State
  Univ., Baton Rouge, LA 70803, gclayton@fenway.phys.lsu.edu}

\begin{abstract} 
We present a sample of 75 extinction curves derived from {\it FUSE}
far-ultraviolet spectra supplemented by existing {\it IUE} spectra.
The extinction curves were created using the standard pair method
based on a new set of dereddened FUSE+IUE comparison stars. Molecular
hydrogen absorption features were removed using individualized
$H_2$ models for each sightline.  The general shape of the FUSE
extinction ($8.4~\micron^{-1} < \lambda^{-1} < 11~\micron^{-1}$) was found to
be broadly consistent with extrapolations from the IUE extinction
($3.3~\micron^{-1} < \lambda^{-1} < 8.6~\micron^{-1}$) curve.  Significant
differences were seen in the 
strength of the far-UV rise and the width of the 2175~\AA\ bump.  All
the FUSE+IUE extinction curves had positive far-UV slopes giving no
indication that the far-UV rise was turning over at the shortest
wavelengths.  The dependence of $A(\lambda)/A(V)$ versus $R(V)^{-1}$
in the far-UV using the sightlines in our sample was found to be stronger than
tentatively indicated by previous work.  We present an updated $R(V)$
dependent relationship for the full UV wavelength range
($3.3~\micron^{-1} \le \lambda^{-1} \le 11~\micron^{-1}$).  Finally, we
searched for discrete absorption features 
in the far-ultraviolet.  We found a 3$\sigma$ upper limit of $\sim$$0.12 A(V)$
on features with a resolution of 250 ($\sim$4~\AA\ width) and
3$\sigma$ upper limits of $\sim$$0.15 A(V)$ for $\lambda^{-1} <
9.6$~\micron$^{-1}$ and $\sim$$0.68 A(V)$ for $\lambda^{-1} >
9.6$~\micron$^{-1}$ on features with a resolution of $10^4$
($\sim$0.1~\AA\ width).
\end{abstract}


\section{Introduction}
\label{sec_intro}

Extinction curves are one of the cornerstones of our understanding of
dust in interstellar space.  Extinction curves probe the sum of the
absorption and scattering of dust and their determination is
observationally straightforward.  Most extinction curves are measured
using the standard pair method \citep{Stecher65, Massa83} where
the measurements of a reddened star are ratioed to those of an
unreddened star of the same spectral type.  The importance of
measuring extinction curves over the widest wavelength range possible
is attested to by their use as a fundamental constraint in models of
dust grains \citep{Weingartner01, Clayton03, Zubko04}.  Measuring
extinction curves is also important for purely empirical reasons in
order to allow for the proper accounting for the effects of
interstellar dust in the study of astrophysical objects.

Structure in interstellar extinction curves gives direct evidence of the
dust grain compositions.  In the ultraviolet, the one very obvious
discrete feature is the 2175~\AA\ extinction bump
\citep{Stecher65, Stecher69}.  The 2175~\AA\ bump has been
observed to have nearly constant central wavelength, a width varying
from 280--660~\AA\ \citep{Fitzpatrick86, Valencic04, Fitzpatrick07},
and has been attributed to small graphite grains
\citep{Stecher65graphite, Draine93}.  There have been no other
discrete features found in the ultraviolet with fairly sensitive
limits placed from searches \citep{Clayton03UVDIBs} for
counterparts of the optical/near-infrared Diffuse Interstellar Bands
\citep[DIBs,][]{Merrill34, Herbig95}.  The other main structure seen in
the ultraviolet extinction curve is the far-ultraviolet (far-UV) rise.
This 
rise is characterized by a constant shape between 1700 and 1175~\AA\
\citep{Fitzpatrick88} and has been modeled as the wing of a feature
like the 2175~\AA\ bump, but centered at 715~\AA\ \citep{Joblin92, Li01}.

The structure in the ultraviolet between 1175 and 3300~\AA\ has been
well studied using spectra taken with the International
Ultraviolet Explorer (IUE) \citep{Fitzpatrick90, Valencic04} and the
Space Telescope Spectrograph on the Hubble Space Telescope
\citep{Clayton03UVDIBs}.  The structure in this wavelength range is
well approximated by the \citet{Fitzpatrick90} (hereafter
FM90) parameterization with 
the curve being the combination of a linear term, a Drude profile,
and a cubic describing the far-UV rise.  In addition, the average
behavior of ultraviolet extinction was found by \citet{Cardelli89}
(hereafter CCM89) to
correlate with $R(V) = A(V)/E(B-V)$ (a rough measure of the average
dust grain size).  More recently, \citet{Fitzpatrick07} have proposed a
revised parameterization that includes an additional parameter and
simplifies the far-UV curvature term to just a quadratic.

\tabletypesize{\scriptsize}
\begin{deluxetable*}{lrrccrccccc}
\tablewidth{0pt}
\tablecaption{Comparison Stars \label{tab_comp_stars}}
\tablehead{ \colhead{name} & \colhead{SpType} & \colhead{V} & 
   \colhead{A(V)} & \colhead{R(V)} & 
   \colhead{C$_1$} & \colhead{C$_2$} & \colhead{C$_3$} & \colhead{C$_4$} &
   \colhead{x$_o$} & \colhead{$\gamma$} }
\startdata
\multicolumn{11}{c}{V, main sequence} \\ \hline
HD091824 &      O6V &  8.14 & $0.60 \pm 0.06$ & $3.10 \pm 0.31$ & $-0.81 \pm  0.41$ & $ 0.90 \pm  0.22$ & $ 3.69 \pm  1.11$ & $ 0.43 \pm  0.06$ & $ 4.62 \pm  0.09$ & $ 1.00 \pm  0.10$ \\ 
HD093028 &      O8V &  8.37 & $0.80 \pm 0.12$ & $3.20 \pm 0.48$ & $ 0.38 \pm  0.19$ & $ 0.61 \pm  0.15$ & $ 2.07 \pm  0.41$ & $ 0.29 \pm  0.06$ & $ 4.62 \pm  0.18$ & $ 0.87 \pm  0.09$ \\ 
HD097471 &      B0V &  9.30 & $0.90 \pm 0.09$ & $3.10 \pm 0.31$ & $-0.93 \pm  0.47$ & $ 0.88 \pm  0.18$ & $ 1.83 \pm  0.55$ & $ 0.28 \pm  0.06$ & $ 4.62 \pm  0.09$ & $ 0.85 \pm  0.09$ \\ 
BD+52$\arcdeg$3210 &      B1V & 10.69 & $0.85 \pm 0.09$ & $3.50 \pm 0.35$ & $-1.67 \pm  0.42$ & $ 0.99 \pm  0.15$ & $ 3.43 \pm  0.69$ & $ 0.25 \pm  0.06$ & $ 4.60 \pm  0.09$ & $ 0.91 \pm  0.09$ \\ 
BD+32$\arcdeg$270 &      B2V & 10.29 & $0.10 \pm 0.04$ & $3.10 \pm 1.24$ & $-0.24 \pm  0.12$ & $ 0.74 \pm  0.37$ & $ 2.00 \pm  1.00$ & $ 0.82 \pm  0.41$ & $ 4.59 \pm  0.46$ & $ 1.00 \pm  0.20$ \\ 
HD051013 &      B3V &  8.81 & $0.10 \pm 0.04$ & $3.10 \pm 1.55$ & $-0.24 \pm  0.12$ & $ 0.74 \pm  0.37$ & $ 2.00 \pm  1.00$ & $ 0.42 \pm  0.21$ & $ 4.59 \pm  0.46$ & $ 1.00 \pm  0.20$ \\ 
HD037332 &      B4V &  7.62 & $0.10 \pm 0.02$ & $3.10 \pm 0.60$ & $-0.24 \pm  0.12$ & $ 0.94 \pm  0.47$ & $ 3.56 \pm  1.78$ & $ 2.00 \pm  0.50$ & $ 4.59 \pm  0.46$ & $ 1.00 \pm  0.20$ \\ 
HD037525 &      B5V &  8.08 & $0.18 \pm 0.09$ & $3.10 \pm 1.24$ & $-0.24 \pm  0.12$ & $ 0.74 \pm  0.37$ & $ 3.56 \pm  1.78$ & $ 0.52 \pm  0.26$ & $ 4.59 \pm  0.18$ & $ 1.00 \pm  0.20$ \\ \hline
\multicolumn{11}{c}{III, giant} \\ \hline
HD116852 &    O9III &  8.47 & $0.63 \pm 0.06$ & $3.01 \pm 0.60$ & $-0.65 \pm  0.33$ & $ 0.70 \pm  0.10$ & $ 1.41 \pm  0.28$ & $ 0.28 \pm  0.07$ & $ 4.56 \pm  0.18$ & $ 0.72 \pm  0.07$ \\ 
HD104705 &    B0III &  7.76 & $1.06 \pm 0.11$ & $3.31 \pm 0.33$ & $ 0.27 \pm  0.13$ & $ 0.49 \pm  0.10$ & $ 2.01 \pm  0.40$ & $ 0.30 \pm  0.03$ & $ 4.56 \pm  0.09$ & $ 0.82 \pm  0.08$ \\ 
HD172140 &  B0.5III &  9.94 & $0.56 \pm 0.11$ & $2.34 \pm 0.47$ & $-0.37 \pm  0.18$ & $ 0.61 \pm  0.15$ & $ 2.58 \pm  0.52$ & $ 0.37 \pm  0.06$ & $ 4.59 \pm  0.09$ & $ 0.89 \pm  0.09$ \\ 
HD114444 &    B2III & 10.31 & $0.51 \pm 0.10$ & $2.65 \pm 0.53$ & $-0.65 \pm  0.32$ & $ 0.40 \pm  0.10$ & $ 2.56 \pm  0.51$ & $ 0.31 \pm  0.08$ & $ 4.61 \pm  0.09$ & $ 0.82 \pm  0.08$ \\ 
HD235874 &    B3III &  9.64 & $0.58 \pm 0.06$ & $2.93 \pm 0.29$ & $-2.63 \pm  0.66$ & $ 1.11 \pm  0.11$ & $ 3.26 \pm  0.65$ & $ 0.26 \pm  0.06$ & $ 4.68 \pm  0.09$ & $ 0.97 \pm  0.10$ \\ \hline 
\multicolumn{11}{c}{I, supergiant} \\ \hline
HD210809 &     O9Ib &  7.54 & $0.98 \pm 0.05$ & $3.10 \pm 0.31$ & $-1.23 \pm  0.31$ & $ 0.98 \pm  0.05$ & $ 2.58 \pm  0.26$ & $ 0.45 \pm  0.02$ & $ 4.60 \pm  0.09$ & $ 1.00 \pm  0.10$ \\ 
HD091983 &   O9.5Ib &  8.58 & $1.10 \pm 0.11$ & $3.10 \pm 0.23$ & $ 0.35 \pm  0.17$ & $ 0.61 \pm  0.09$ & $ 2.50 \pm  0.50$ & $ 0.40 \pm  0.04$ & $ 4.61 \pm  0.09$ & $ 0.98 \pm  0.10$ \\ 
HD094493 &  B0.5Iab &  7.27 & $0.70 \pm 0.04$ & $3.30 \pm 0.33$ & $-1.02 \pm  0.51$ & $ 0.95 \pm  0.19$ & $ 3.14 \pm  0.63$ & $ 0.56 \pm  0.08$ & $ 4.60 \pm  0.09$ & $ 0.93 \pm  0.09$ \\ 
HD100276 &     B1Ib &  7.22 & $0.83 \pm 0.08$ & $3.50 \pm 0.35$ & $-1.48 \pm  0.37$ & $ 0.92 \pm  0.09$ & $ 3.30 \pm  0.66$ & $ 0.26 \pm  0.06$ & $ 4.59 \pm  0.09$ & $ 0.95 \pm  0.10$ \\ 
HD075309 &     B2Ib &  7.85 & $0.64 \pm 0.06$ & $3.53 \pm 0.71$ & $ 0.39 \pm  0.20$ & $ 0.53 \pm  0.13$ & $ 2.59 \pm  0.52$ & $ 0.30 \pm  0.08$ & $ 4.60 \pm  0.09$ & $ 0.76 \pm  0.08$ 
\enddata
\end{deluxetable*}

The dust extinction in the far-UV region between 912
and 1190~\AA\ has only been studied for a handful of Milky Way sightlines
\citep[see][for a review]{Sofia05}.  Our understanding of the far-UV
extinction and the nature of the dust materials responsible for it is,
necessarily, tentative.  The many spectra of hot stars taken with the
Far-Ultraviolet Spectroscopic Explorer \citep[FUSE,][]{Moos00} allow
for an extensive study of the extinction curve in the far-UV region
and motivate this analysis.  The aim is to
refine our understanding of the far-UV extinction using a
statistically significant sample of FUSE extinction curves.
Specifically, we will investigate how well the FM90,
CCM89, and \citet{Fitzpatrick07} parameterizations
extrapolate to the far-UV and refine them if necessary.  Also,
we will see if the far-UV region contains any significant structure in
addition to the well-known far-UV rise.

This study is the third paper in a series of papers investigating FUSE
far-UV extinction curves.  The first paper in this series
\citep{Sofia05} presented a preliminary study of far-UV extinction
curves along nine Milky Way sightlines.  The second paper in this series
\citep{Cartledge05} presented a study of nine far-UV extinction curves
in the Magellanic Clouds.  This third paper presents far-UV extinction
curves for 75 sightlines and benefits from a more mature FUSE
calibration, a better understanding of correcting for the $H_2$
absorption, and a more complete set of FUSE comparison stars.  The
fourth paper in this series \citet{Cartledge09} will study
the correlations between the gas properties (e.g., $H_2$, \& \ion{H}{1}) and
dust extinction.

\section{Data}
\label{sec_data}

The sample of stars for this paper was determined by searching the
FUSE archive for stars from the \citet{Valencic04} (hereafter VGC04)
sample of IUE 
extinction curves that had good quality FUSE spectra.  IUE extinction
curves are required because the large scale
extinction structure in the FUSE wavelength range (far-UV rise) starts
in the IUE spectral range around 1700~\AA\ \citep{Fitzpatrick88}.  As
a side benefit, the existence of IUE extinction curves for all the
sightlines studied in this paper ensures that these sightlines are
suitable for extinction curve determinations.  The final sample for
this paper includes 75 sightlines.

The FUSE observations (912 -- 1190~\AA) were extracted from the
archive and reduced using CalFUSE v3.0.  The observations
used for each star are given in \citet{Cartledge09}.  The next step
was to generate corrections for the copious $H_2$ absorption seen
throughout the FUSE spectral region.  This was done by determining the
physical parameters of the $H_2$ absorbing gas and generating a model
spectrum of the $H_2$ absorption.  The details of the $H_2$ and \ion{H}{1}
modeling are given in \citet{Cartledge09}.  When correcting a
FUSE spectrum for $H_2$ and \ion{H}{1} absorption, any point which is absorbed
by more than 70\% was excluded from use in generating an
extinction curve, to minimize the effects of the uncertainties in the
$H_2$ and \ion{H}{1} corrections on the resulting extinction curves.
The residual effects of the $H_2$ and \ion{H}{1} absorption on the
final extinction curves are discussed in \S\ref{sec_structure}.

The complimentary IUE spectra (1150 -- 3225~\AA) were taken from the
IUE archive following VCG04.  Optical and near-infrared
photometry was accumulated from the literature also following
VCG04.  The spectral overlap between the IUE and FUSE observations
was used to derive a multiplicative correction to the FUSE spectra to
put them on the same flux scale as the IUE spectra.  The corrections
are generally small with maximum corrections up to 25\%.  Finally, the
FUSE and IUE spectra were rebinned to a common resolution of 250 which
is sufficient for most extinction curve work.

\subsection{New Comparison Stars}
\label{sec_comp_stars}

Calculating an extinction curve for the sightline towards a reddened star
requires an unreddened comparison star of the same spectral type.  For IUE
observations, there exists a good set of bright, lightly reddened comparison
stars which have been carefully dereddened \citep{Cardelli92}.
Ideally, FUSE observations of these same IUE comparison stars would be
used to extend their spectra to far-UV wavelengths.  Unfortunately, many
of the IUE comparison stars are too bright to be observable with
FUSE.  A new set of fainter comparison stars that are observable with
FUSE is required.

\begin{figure*}[htbp]
\epsscale{1.2}
\plotone{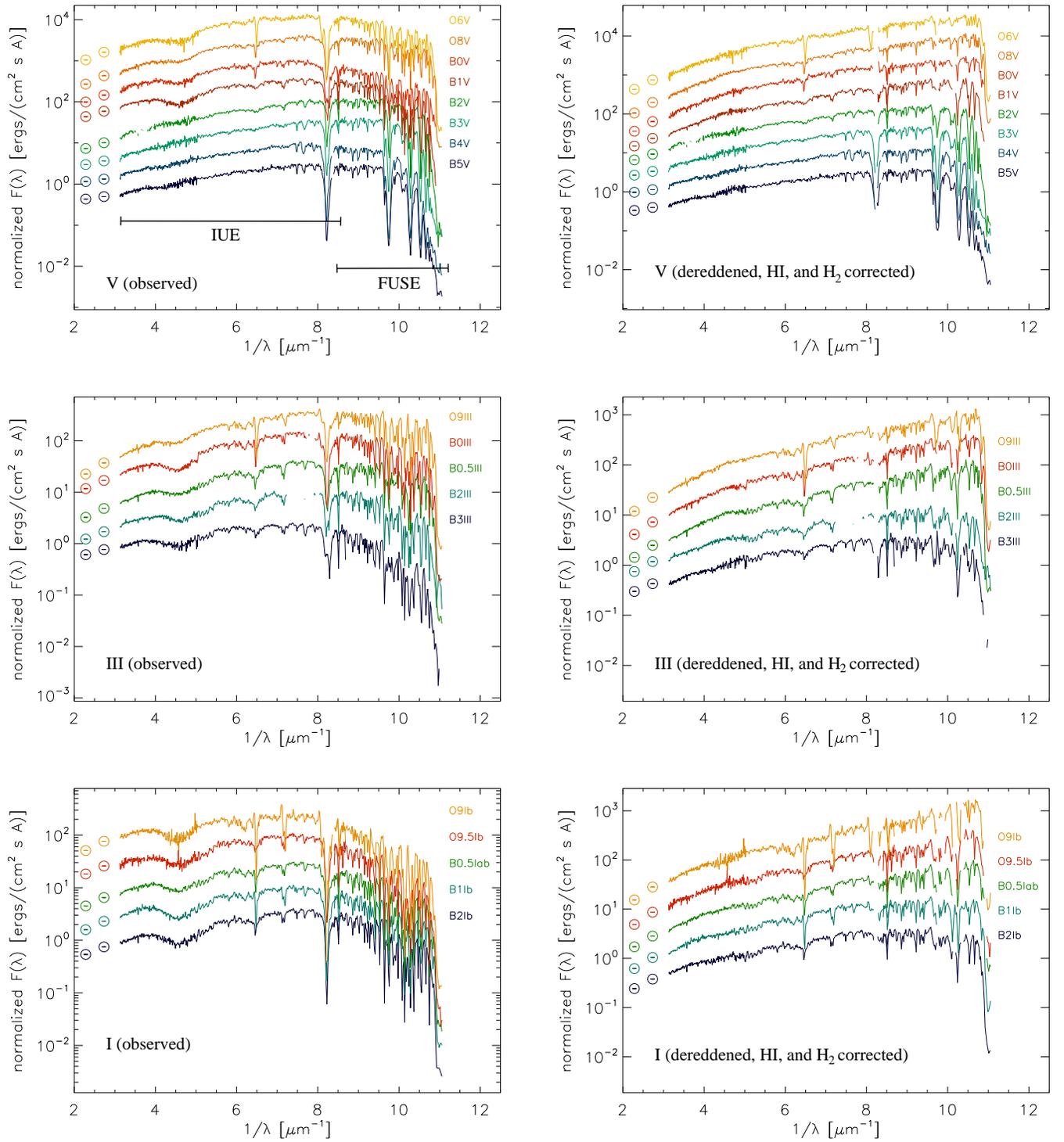}
\caption{The FUSE and IUE spectra of the comparison stars are plotted.
All the spectra are plotted at a common spectral resolution of 250.
The plots on the left give the observed spectra and the same spectra
which have been dereddened and
corrected for \ion{H}{1} and H$_2$ absorption are on the right.  The
spectra are normalized to one between 4 and 5 \micron$^{-1}$ and
offset from each other by 0.5 in log space.
\label{fig_standards}}
\end{figure*}

We have generated the FUSE set of comparison stars by picking the least
reddened star of each spectral type that has both FUSE and IUE
observations.  The new comparison star sample is listed in
Table~\ref{tab_comp_stars} along with the parameters used to deredden
their IUE and FUSE spectra.  The $H_2$ and \ion{H}{1} model parameters used to remove
the affects of $H_2$ absorption are given by \citep{Cartledge09}.
Given the FUSE brightness limit, these 
stars are necessarily more distant and reddened than the IUE comparison star
sample.  We have carefully dereddened these new comparison stars
following the methods of \citet{Cardelli92}.  The dereddening was done
starting with $E(B-V)$, $A(V)$, and \citet{Fitzpatrick90}  parameters taken from 
VCG04.  Thus, these new comparison stars are directly
bootstrapped off of the existing IUE comparison stars.  The use of FM
parameters determined in the IUE spectral range to deredden FUSE
spectra is supported by earlier work on FUSE extinction curves showing
that the extrapolation of the FM fits to the FUSE wavelength range is
reasonable \citep{Sofia05}.  We have checked that this extrapolation
is not affecting our results in \S\ref{sec_rv_rel}.  As was done by 
\citet{Cardelli92}, the dereddening parameters were then manually
tweaked to produce dereddened spectra that lacked obvious dust
extinction and had a smooth progression between spectral types.  This
is illustrated in Fig.~\ref{fig_standards} where the observed and
dereddened spectra for each class of standards (main sequence, giants,
and supergiants) are shown.  The uncertainties on the dereddening
parameters was estimated by changing each parameter until it was
clearly incorrect.  The resulting uncertainties are given in
Table~\ref{tab_comp_stars} and are assumed to be one sigma
uncertainties to be conservative.

An alternative to using observed comparison stars is to use stellar
atmosphere models.  This is an approach that has been explored by
\citet{Fitzpatrick99, Fitzpatrick05} and found to produce similar
results to using observed comparison stars.  The strength of using
model atmospheres is that it removes the need to deredden observed
spectra and should provide closer spectral matches.  The weakness is
that it relies on the accuracy of the stellar atmosphere models and
the absolute calibration of the spectra.  Observed comparison stars
only rely on the relative calibration of the spectra and, by
definition, include the same physics as the reddened stars.  The use
of observed comparison stars does inject some additional uncertainty
due to the lack of ``perfect'' matches to the reddened star spectral
types.  In the end, using observed or model comparison stars represent
complimentary approaches, each with strengths and weaknesses. The fact
that they produce similar results provides strong evidence that
the resulting extinction curves derived from either method are
correct.

\subsection{Extinction Curve Calculation}

The extinction curves were calculated using the standard pair method
\citep{Stecher65, Massa83}.  Basically, the ratio of the fluxes of the
reddened and comparison stars (both with the same spectral type) gives
a direct measurement of the dust extinction towards the reddened star.
Ideally, the distances to the reddened and comparison stars would be
known to high accuracy, allowing for an absolute measurement of the
dust extinction.  Unfortunately, the distances to the hot, early-type
stars used for ultraviolet dust extinction curve measurements are
rarely, if ever, known to high enough accuracy.  It is possible then
to make a differential measurement of the difference between
the colors of reddened and comparison stars.  The basic measurement is
\begin{equation}
E(\lambda - V) = m(\lambda - V)_r - m(\lambda - V)_c
\end{equation}
where $r$ and $c$ subscripts refer to the reddened and comparison
stars, respectively.  This measurement is the unnormalized extinction
curve.  The use of colors w.r.t. the V band removes the unknown or
uncertain distance to each of the stars from the measurement.  In
order to make comparisons between extinction curves measured along
different sightlines, the curve needs to be normalized to a
measurement of the total column of dust along the sightline.  The most
often used normalization is to divide $E(\lambda-V)$ by $E(B-V)$.  Yet
a more direct measure of the dust properties is $A(\lambda)/A(V)$.

There are two avenues to determining $A(\lambda)/A(V)$ from the basic
$E(\lambda-V)$ measurement.  The usual method is to determine the
conversion factor $R(V)$ [$= A(V)/E(B-V)$] by extrapolating the
$E(\lambda-V)/E(B-V)$ curve at JHK wavelengths to infinite wavelength
using an assumed form of the $A(\lambda)/A(V)$ curve at wavelengths $>
1$~\micron.  The conversion is then
\begin{equation}
\frac{A(\lambda)}{A(V)} = \frac{1}{R(V)}
  \left[ \frac{E(\lambda-V)}{E(B-V)} \right] + 1
\end{equation}
A more direct way is to extrapolate the $E(\lambda-V)$ curve to
infinite wavelength using the same assumptions [JHK wavelengths and an
assumed $A(\lambda)/A(V)$ curve] to determine $A(V)$.  The conversion
is then
\begin{equation}
\frac{A(\lambda)}{A(V)} = \frac{E(\lambda-V)}{A(V)} + 1.
\end{equation}
Functionally, both methods of determining $A(\lambda)/A(V)$ are
equivalent.

\begin{figure*}[tbp]
\epsscale{1.2}
\plotone{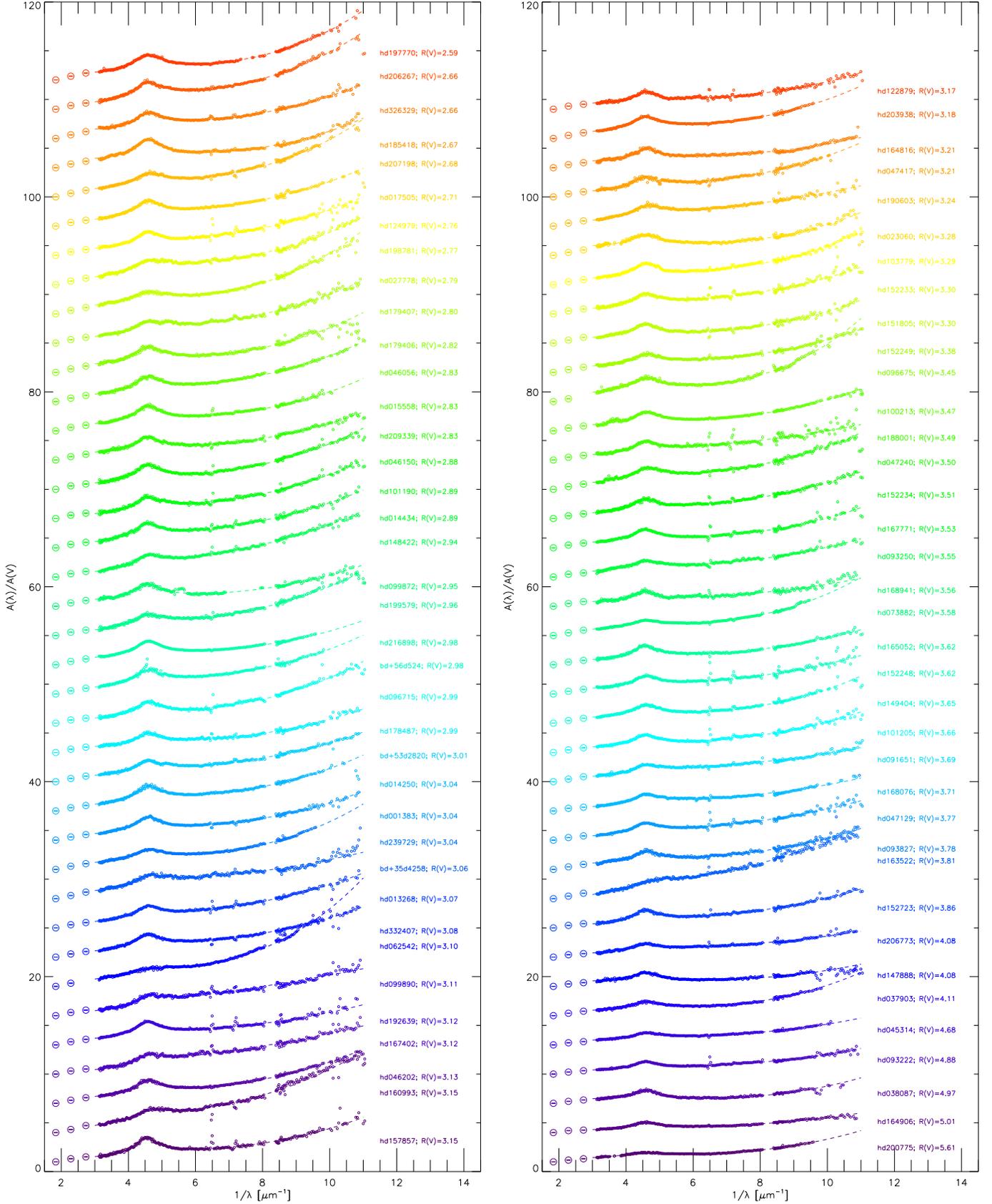}
\caption{The extinction curves are plotted for the entire sample
along with the FM90 fit to 
each curve.  The curves are plotted sorted by measured $R(V)$ values
and offset by a constant value from each other.  The reddened star
name and measured $R(V)$ value is given to the right of each curve.
\label{fig_ext_curves1}}
\end{figure*}

\tabletypesize{\scriptsize}
\begin{deluxetable*}{rrrrrrr}
\tablewidth{0pt}
\tablecaption{Extinction Curve Details \label{tab_ext_info}}
\tablehead{ \colhead{name} & \colhead{SpType} & \colhead{comparison} & 
   \colhead{V} & \colhead{A(V)\tablenotemark{a}} & \colhead{E(B-V)\tablenotemark{a}} & \colhead{R(V)\tablenotemark{a}} }
\startdata
BD+35$\arcdeg$4258 &    B0.5V &   HD097471 &  9.41 & $ 0.95 \pm  0.04 \pm  0.03$ & $ 0.31 \pm 0.053 \pm 0.037$ & $ 3.06 \pm 0.380 \pm 0.011$ \\
BD+53$\arcdeg$820 &     B0IV &   HD104705 &  9.95 & $ 1.20 \pm  0.04 \pm  0.03$ & $ 0.40 \pm 0.047 \pm 0.035$ & $ 3.01 \pm 0.249 \pm 0.012$ \\
 BD+56$\arcdeg$524 &      B1V & BD+52$\arcdeg$3210 &  9.75 & $ 1.80 \pm  0.04 \pm  0.02$ & $ 0.60 \pm 0.044 \pm 0.023$ & $ 2.98 \pm 0.191 \pm 0.009$ \\
  HD001383 &     B1II &   HD100276 &  7.63 & $ 1.36 \pm  0.04 \pm  0.02$ & $ 0.45 \pm 0.035 \pm 0.029$ & $ 3.04 \pm 0.151 \pm 0.009$ \\
  HD013268 &      O8V &   HD093028 &  8.18 & $ 1.35 \pm  0.04 \pm  0.03$ & $ 0.44 \pm 0.052 \pm 0.042$ & $ 3.07 \pm 0.231 \pm 0.014$ \\
  HD014250 &     B1IV & BD+52$\arcdeg$3210 &  8.96 & $ 1.77 \pm  0.04 \pm  0.02$ & $ 0.58 \pm 0.044 \pm 0.024$ & $ 3.04 \pm 0.203 \pm 0.009$ \\
  HD014434 &    O6.5V &   HD091824 &  8.50 & $ 1.22 \pm  0.06 \pm  0.02$ & $ 0.42 \pm 0.044 \pm 0.021$ & $ 2.89 \pm 0.293 \pm 0.007$ \\
  HD015558 &      O5V &   HD091824 &  7.81 & $ 2.22 \pm  0.05 \pm  0.02$ & $ 0.78 \pm 0.041 \pm 0.013$ & $ 2.83 \pm 0.153 \pm 0.007$ \\
  HD017505 &      O6V &   HD091824 &  7.06 & $ 1.80 \pm  0.06 \pm  0.02$ & $ 0.66 \pm 0.041 \pm 0.014$ & $ 2.71 \pm 0.179 \pm 0.007$ \\
  HD023060 &      B2V &  BD+32$\arcdeg$270 &  7.48 & $ 0.99 \pm  0.04 \pm  0.01$ & $ 0.30 \pm 0.056 \pm 0.028$ & $ 3.28 \pm 0.542 \pm 0.004$ \\
  HD027778 &      B3V &   HD051013 &  6.36 & $ 1.09 \pm  0.03 \pm  0.01$ & $ 0.39 \pm 0.057 \pm 0.023$ & $ 2.79 \pm 0.382 \pm 0.005$ \\
  HD037903 &    B1.5V & BD+52$\arcdeg$3210 &  7.83 & $ 1.49 \pm  0.04 \pm  0.02$ & $ 0.36 \pm 0.054 \pm 0.038$ & $ 4.11 \pm 0.432 \pm 0.009$ \\
  HD038087 &      B5V &   HD037525 &  8.30 & $ 1.33 \pm  0.04 \pm  0.03$ & $ 0.27 \pm 0.082 \pm 0.079$ & $ 4.97 \pm 0.427 \pm 0.009$ \\
  HD045314 &      O9V &   HD093028 &  6.64 & $ 2.15 \pm  0.04 \pm  0.03$ & $ 0.46 \pm 0.062 \pm 0.054$ & $ 4.68 \pm 0.328 \pm 0.013$ \\
  HD046056 &      O8V &   HD093028 &  8.15 & $ 1.41 \pm  0.05 \pm  0.03$ & $ 0.50 \pm 0.048 \pm 0.036$ & $ 2.83 \pm 0.190 \pm 0.013$ \\
  HD046150 &      O6V &   HD091824 &  6.76 & $ 1.13 \pm  0.05 \pm  0.02$ & $ 0.39 \pm 0.045 \pm 0.022$ & $ 2.88 \pm 0.309 \pm 0.007$ \\
  HD046202 &      O9V &   HD093028 &  8.18 & $ 1.53 \pm  0.05 \pm  0.03$ & $ 0.49 \pm 0.048 \pm 0.037$ & $ 3.13 \pm 0.213 \pm 0.014$ \\
  HD047129 &      O8V &   HD093028 &  6.08 & $ 1.39 \pm  0.05 \pm  0.04$ & $ 0.37 \pm 0.060 \pm 0.051$ & $ 3.77 \pm 0.332 \pm 0.014$ \\
  HD047240 &     B1Ib &   HD100276 &  6.15 & $ 1.14 \pm  0.04 \pm  0.02$ & $ 0.33 \pm 0.042 \pm 0.037$ & $ 3.50 \pm 0.232 \pm 0.009$ \\
  HD047417 &     B0IV &   HD104705 &  6.97 & $ 0.99 \pm  0.04 \pm  0.03$ & $ 0.31 \pm 0.052 \pm 0.041$ & $ 3.21 \pm 0.340 \pm 0.012$ \\
  HD062542 &      B3V &   HD051013 &  8.04 & $ 1.16 \pm  0.04 \pm  0.01$ & $ 0.37 \pm 0.063 \pm 0.026$ & $ 3.10 \pm 0.491 \pm 0.005$ \\
  HD073882 &    O9III &   HD116852 &  7.21 & $ 2.46 \pm  0.03 \pm  0.02$ & $ 0.69 \pm 0.042 \pm 0.037$ & $ 3.58 \pm 0.112 \pm 0.011$ \\
  HD091651 &      O9V &   HD093028 &  8.84 & $ 1.07 \pm  0.05 \pm  0.04$ & $ 0.29 \pm 0.075 \pm 0.060$ & $ 3.69 \pm 0.594 \pm 0.014$ \\
  HD093222 &      O8V &   HD093028 &  8.11 & $ 1.76 \pm  0.05 \pm  0.04$ & $ 0.36 \pm 0.077 \pm 0.066$ & $ 4.88 \pm 0.550 \pm 0.014$ \\
  HD093250 &      O6V &   HD091824 &  7.37 & $ 1.54 \pm  0.05 \pm  0.02$ & $ 0.43 \pm 0.046 \pm 0.025$ & $ 3.55 \pm 0.336 \pm 0.008$ \\
  HD093827 &     B2Ib &   HD075309 &  9.31 & $ 0.80 \pm  0.04 \pm  0.02$ & $ 0.21 \pm 0.091 \pm 0.083$ & $ 3.78 \pm 0.709 \pm 0.010$ \\
  HD096675 &     B6IV &   HD037525 &  7.69 & $ 0.99 \pm  0.04 \pm  0.03$ & $ 0.29 \pm 0.058 \pm 0.054$ & $ 3.45 \pm 0.284 \pm 0.009$ \\
  HD096715 &      O5V &   HD091824 &  8.25 & $ 1.12 \pm  0.05 \pm  0.02$ & $ 0.37 \pm 0.046 \pm 0.024$ & $ 2.99 \pm 0.338 \pm 0.008$ \\
  HD099872 &      B3V &   HD051013 &  6.09 & $ 1.07 \pm  0.03 \pm  0.01$ & $ 0.36 \pm 0.058 \pm 0.026$ & $ 2.95 \pm 0.435 \pm 0.005$ \\
  HD099890 &    B0.5V &   HD097471 &  8.28 & $ 0.75 \pm  0.04 \pm  0.03$ & $ 0.24 \pm 0.067 \pm 0.043$ & $ 3.11 \pm 0.675 \pm 0.011$ \\
  HD100213 &      O8V &   HD093028 &  8.38 & $ 1.28 \pm  0.05 \pm  0.04$ & $ 0.37 \pm 0.059 \pm 0.050$ & $ 3.47 \pm 0.307 \pm 0.014$ \\
  HD101190 &      O7V &   HD091824 &  7.27 & $ 0.90 \pm  0.05 \pm  0.02$ & $ 0.31 \pm 0.047 \pm 0.026$ & $ 2.89 \pm 0.390 \pm 0.007$ \\
  HD101205 &      O8V &   HD093028 &  6.47 & $ 1.28 \pm  0.05 \pm  0.03$ & $ 0.35 \pm 0.066 \pm 0.052$ & $ 3.66 \pm 0.429 \pm 0.014$ \\
  HD103779 &   B0.5II &   HD094493 &  7.21 & $ 0.66 \pm  0.03 \pm  0.01$ & $ 0.20 \pm 0.045 \pm 0.041$ & $ 3.29 \pm 0.356 \pm 0.006$ \\
  HD122879 &     B0Ia &   HD091983 &  6.41 & $ 1.41 \pm  0.04 \pm  0.03$ & $ 0.44 \pm 0.036 \pm 0.025$ & $ 3.17 \pm 0.199 \pm 0.012$ \\
  HD124979 &      O8V &   HD093028 &  8.50 & $ 1.19 \pm  0.05 \pm  0.04$ & $ 0.43 \pm 0.049 \pm 0.038$ & $ 2.76 \pm 0.204 \pm 0.014$ \\
  HD147888 &      B4V &   HD037332 &  6.74 & $ 1.97 \pm  0.03 \pm  0.01$ & $ 0.48 \pm 0.023 \pm 0.011$ & $ 4.08 \pm 0.184 \pm 0.002$ \\
  HD148422 &     B1Ia &   HD100276 &  8.65 & $ 0.84 \pm  0.04 \pm  0.02$ & $ 0.29 \pm 0.042 \pm 0.036$ & $ 2.94 \pm 0.229 \pm 0.009$ \\
  HD149404 &     O9Ia &   HD210809 &  5.48 & $ 2.28 \pm  0.06 \pm  0.01$ & $ 0.63 \pm 0.043 \pm 0.029$ & $ 3.65 \pm 0.204 \pm 0.009$ \\
  HD151805 &     B1Ib &   HD100276 &  8.91 & $ 1.11 \pm  0.04 \pm  0.02$ & $ 0.34 \pm 0.044 \pm 0.036$ & $ 3.30 \pm 0.275 \pm 0.009$ \\
  HD152233 &      O6V &   HD091824 &  6.59 & $ 1.30 \pm  0.06 \pm  0.02$ & $ 0.39 \pm 0.052 \pm 0.025$ & $ 3.30 \pm 0.407 \pm 0.007$ \\
  HD152234 &   B0.5Ia &   HD094493 &  5.45 & $ 1.38 \pm  0.04 \pm  0.01$ & $ 0.39 \pm 0.044 \pm 0.031$ & $ 3.51 \pm 0.303 \pm 0.006$ \\
  HD152248 &      O8V &   HD093028 &  6.10 & $ 1.66 \pm  0.05 \pm  0.04$ & $ 0.46 \pm 0.055 \pm 0.045$ & $ 3.62 \pm 0.260 \pm 0.014$ \\
  HD152249 &     O9Ib &   HD210809 &  6.45 & $ 1.58 \pm  0.03 \pm  0.02$ & $ 0.47 \pm 0.079 \pm 0.032$ & $ 3.38 \pm 0.527 \pm 0.009$ \\
  HD152723 &      O7V &   HD091824 &  7.31 & $ 1.40 \pm  0.06 \pm  0.02$ & $ 0.36 \pm 0.055 \pm 0.030$ & $ 3.86 \pm 0.508 \pm 0.007$ \\
  HD157857 &      O7V &   HD091824 &  7.78 & $ 1.37 \pm  0.05 \pm  0.02$ & $ 0.43 \pm 0.045 \pm 0.023$ & $ 3.15 \pm 0.303 \pm 0.007$ \\
  HD160993 &    B1Iab &   HD100276 &  7.71 & $ 0.65 \pm  0.04 \pm  0.02$ & $ 0.21 \pm 0.060 \pm 0.046$ & $ 3.15 \pm 0.593 \pm 0.009$ \\
  HD163522 &     B1Ia &   HD100276 &  8.43 & $ 0.71 \pm  0.04 \pm  0.02$ & $ 0.19 \pm 0.064 \pm 0.053$ & $ 3.81 \pm 0.785 \pm 0.009$ \\
  HD164816 &      B0V &   HD097471 &  7.11 & $ 0.99 \pm  0.04 \pm  0.03$ & $ 0.31 \pm 0.059 \pm 0.038$ & $ 3.21 \pm 0.474 \pm 0.010$ \\
  HD164906 &     B1IV & BD+52$\arcdeg$3210 &  7.47 & $ 2.17 \pm  0.04 \pm  0.02$ & $ 0.43 \pm 0.054 \pm 0.040$ & $ 5.01 \pm 0.440 \pm 0.008$ \\
  HD165052 &    O6.5V &   HD091824 &  6.87 & $ 1.25 \pm  0.06 \pm  0.02$ & $ 0.35 \pm 0.064 \pm 0.029$ & $ 3.62 \pm 0.616 \pm 0.007$ \\
  HD167402 &    O9.5V &   HD097471 &  9.03 & $ 0.88 \pm  0.04 \pm  0.03$ & $ 0.28 \pm 0.058 \pm 0.041$ & $ 3.12 \pm 0.471 \pm 0.010$ \\
  HD167771 &      O8V &   HD093028 &  6.54 & $ 1.48 \pm  0.05 \pm  0.04$ & $ 0.42 \pm 0.060 \pm 0.048$ & $ 3.53 \pm 0.311 \pm 0.013$ \\
  HD168076 &      O5V &   HD091824 &  8.24 & $ 2.61 \pm  0.05 \pm  0.02$ & $ 0.70 \pm 0.043 \pm 0.019$ & $ 3.71 \pm 0.216 \pm 0.007$ \\
  HD168941 &   O9.5II &   HD091983 &  9.38 & $ 1.23 \pm  0.04 \pm  0.03$ & $ 0.34 \pm 0.042 \pm 0.032$ & $ 3.56 \pm 0.284 \pm 0.011$ \\
  HD178487 &     B0Ia &   HD091983 &  8.66 & $ 1.42 \pm  0.04 \pm  0.03$ & $ 0.47 \pm 0.032 \pm 0.025$ & $ 2.99 \pm 0.141 \pm 0.011$ \\
  HD179406 &      B3V &   HD051013 &  5.33 & $ 0.94 \pm  0.03 \pm  0.01$ & $ 0.33 \pm 0.060 \pm 0.030$ & $ 2.82 \pm 0.456 \pm 0.004$ \\
  HD179407 &     B0II &   HD104705 &  9.41 & $ 1.09 \pm  0.04 \pm  0.03$ & $ 0.39 \pm 0.049 \pm 0.037$ & $ 2.80 \pm 0.241 \pm 0.011$ \\
  HD185418 &    B0.5V &   HD097471 &  7.45 & $ 1.39 \pm  0.04 \pm  0.03$ & $ 0.52 \pm 0.046 \pm 0.027$ & $ 2.67 \pm 0.200 \pm 0.010$ \\
  HD188001 &      O8V &   HD093028 &  6.22 & $ 1.12 \pm  0.05 \pm  0.03$ & $ 0.32 \pm 0.066 \pm 0.057$ & $ 3.49 \pm 0.358 \pm 0.013$ \\
  HD190603 &   B1.5Ia &   HD100276 &  5.64 & $ 2.35 \pm  0.29 \pm  0.02$ & $ 0.73 \pm 0.034 \pm 0.022$ & $ 3.24 \pm 0.410 \pm 0.008$ \\
  HD192639 &      O8V &   HD093028 &  7.11 & $ 2.06 \pm  0.04 \pm  0.03$ & $ 0.66 \pm 0.047 \pm 0.034$ & $ 3.12 \pm 0.155 \pm 0.012$ \\
  HD197770 &    B2III &   HD114444 &  6.32 & $ 1.43 \pm  0.04 \pm  0.03$ & $ 0.55 \pm 0.045 \pm 0.034$ & $ 2.59 \pm 0.145 \pm 0.012$ \\
  HD198781 &    B0.5V &   HD097471 &  6.45 & $ 1.03 \pm  0.04 \pm  0.03$ & $ 0.37 \pm 0.050 \pm 0.033$ & $ 2.77 \pm 0.292 \pm 0.010$ \\
  HD199579 &      O6V &   HD091824 &  5.96 & $ 0.93 \pm  0.05 \pm  0.02$ & $ 0.31 \pm 0.048 \pm 0.028$ & $ 2.96 \pm 0.400 \pm 0.007$ \\
  HD200775 &     B2Ve &  BD+32$\arcdeg$270 &  7.39 & $ 3.21 \pm  0.04 \pm  0.01$ & $ 0.57 \pm 0.074 \pm 0.027$ & $ 5.61 \pm 0.680 \pm 0.004$ \\
  HD203938 &   B0.5IV &   HD172140 &  7.08 & $ 2.35 \pm  0.04 \pm  0.03$ & $ 0.74 \pm 0.040 \pm 0.035$ & $ 3.18 \pm 0.092 \pm 0.016$ \\
  HD206267 &      O6V &   HD091824 &  5.62 & $ 1.26 \pm  0.05 \pm  0.02$ & $ 0.47 \pm 0.044 \pm 0.020$ & $ 2.66 \pm 0.242 \pm 0.007$ \\
  HD206773 &      B0V &   HD097471 &  6.87 & $ 2.16 \pm  0.04 \pm  0.03$ & $ 0.53 \pm 0.050 \pm 0.033$ & $ 4.08 \pm 0.293 \pm 0.010$ \\
  HD207198 &     O9II &   HD210809 &  5.94 & $ 1.54 \pm  0.03 \pm  0.01$ & $ 0.58 \pm 0.032 \pm 0.026$ & $ 2.68 \pm 0.106 \pm 0.009$ \\
  HD209339 &     B0IV &   HD104705 &  6.65 & $ 1.05 \pm  0.04 \pm  0.03$ & $ 0.37 \pm 0.054 \pm 0.037$ & $ 2.83 \pm 0.320 \pm 0.011$ \\
  HD216898 &    O8.5V &   HD093028 &  8.00 & $ 2.50 \pm  0.05 \pm  0.03$ & $ 0.84 \pm 0.041 \pm 0.027$ & $ 2.98 \pm 0.118 \pm 0.013$ \\
  HD239729 &      B0V &   HD097471 &  8.35 & $ 2.01 \pm  0.04 \pm  0.03$ & $ 0.66 \pm 0.044 \pm 0.024$ & $ 3.04 \pm 0.178 \pm 0.010$ \\
  HD326329 &      O9V &   HD093028 &  8.60 & $ 1.28 \pm  0.09 \pm  0.03$ & $ 0.48 \pm 0.163 \pm 0.038$ & $ 2.66 \pm 0.896 \pm 0.013$ \\
  HD332407 &     B1Ib &   HD100276 &  8.50 & $ 1.19 \pm  0.04 \pm  0.02$ & $ 0.39 \pm 0.039 \pm 0.033$ & $ 3.08 \pm 0.179 \pm 0.009$
\enddata
\tablenotetext{a}{The quantitities are given as value $\pm$ random uncertainty $\pm$ systematic uncertainty.}
\end{deluxetable*}

The use of $A(\lambda)/A(V)$ extinction curves is preferred
over performing our analysis on $E(\lambda-V)/E(B-V)$ extinction
curves as $A(\lambda)/A(V)$ is the more fundamental measurement of the
properties of dust grains.  In addition, the $A(\lambda)/A(V)$ curve
is less affected by systematic uncertainties introduced by the
normalization.  This is not surprising as the fractional
uncertainty on $A(V)$ is lower (on average 3x) than the fractional
uncertainty on $E(B-V)$.  This is due to the quality of the 2MASS
observations that provide the JHK photometry and the fact that the
A(V) measurement is based on (effectively) the average of the JHK
measurements of the extinction curve.  Of course, the
$A(\lambda)/A(V)$ curve does include the assumption that the $>
1$~\micron\ curve can be extrapolated accurately to derive $A(V)$.
This appears to be a reasonable assumption for the diffuse ISM
\citep{Martin90}, 
but may not be for sightlines that probe dense ISM regions
\citep{Indebetouw05, Flaherty07}.  Most (if not all) of the sightlines
in this paper qualify as diffuse sightlines.  We
use the \citet{Rieke89} infrared extinction curve to extrapolate the
JHK $E(\lambda-V)$ curves to infinite wavelength to derive $A(V)$
values.  The \citet{Rieke89} work refines the results of
\citet{Rieke85}.

One useful benefit of directly determining $A(\lambda)/A(V)$ from
$E(\lambda-V)$ is that this simplifies the calculation of the
uncertainty on the normalized extinction curve.
Previously, we would calculate
$E(\lambda-V)/E(B-V)$ curve, determine $R(V)$, and then renormalize the
curve to $A(\lambda)/A(V)$ \citep{Gordon98}.  Propagating uncertainties
using these steps adds the uncertainty in $E(B-V)$ twice in the final
$A(\lambda)/A(V)$ curve when, in fact, the $E(B-V)$ uncertainty does not
contribute to the final uncertainties in $A(\lambda)/A(V)$ at all.
This resulted in the uncertainties we have quoted in some of our previous
papers \citep{Gordon03, Valencic04} being overestimates.
The correct $A(\lambda)/A(V)$ uncertainties are
\begin{eqnarray}
\sigma \left[ \frac{A(\lambda)}{A(V)} \right]^2 & = & 
     \left( \frac{A(\lambda)}{A(V)} \right)^2
     \left( \left\{ \frac{\sigma [E(\lambda-V)]}{E(\lambda-V)}
            \right\}^2 + 
            \left\{ \frac{\sigma [A(V)]}{A(V)} 
            \right\}^2 \right) \nonumber \\
  & + & \sigma_d(\lambda)^2
\end{eqnarray}
where
\begin{equation}
\sigma [ E(\lambda-V) ] = \sqrt{ \sigma [m(\lambda)_r]^2 +
                                 \sigma [V_r]^2 +
                                 \sigma [m(\lambda)_c]^2 +
                                 \sigma [V_c]^2 }
\end{equation}
and $\sigma_d(\lambda)$ is the uncertainty due to dereddening the
comparison stars (\S\ref{sec_comp_stars}).  The value of
$\sigma_d(\lambda)$ for each curve was determined using a Monte Carlo
method where the dereddening is done using parameters picked from the
distribution allowed by the uncertainties on the dereddening
parameters and computing the resulting uncertainty on each
$A(\lambda)/A(V)$ point.  As can be seen from these equations, the
$A(\lambda)/A(V)$ uncertainties can be divided into random (varying
from point to point) and systematic (causing the entire curve to move
up or down) terms.  The random uncertainties are due to the
$\sigma[m(\lambda)_r]$ and $\sigma[m(\lambda)_r]$ terms.  The
systematic uncertainties are due to the $\sigma[V_r]$, $\sigma[V_c]$,
$\sigma[A(V)]$, and $\sigma_d(\lambda)$ terms.

\begin{figure}[tbp]
\epsscale{1.2}
\plotone{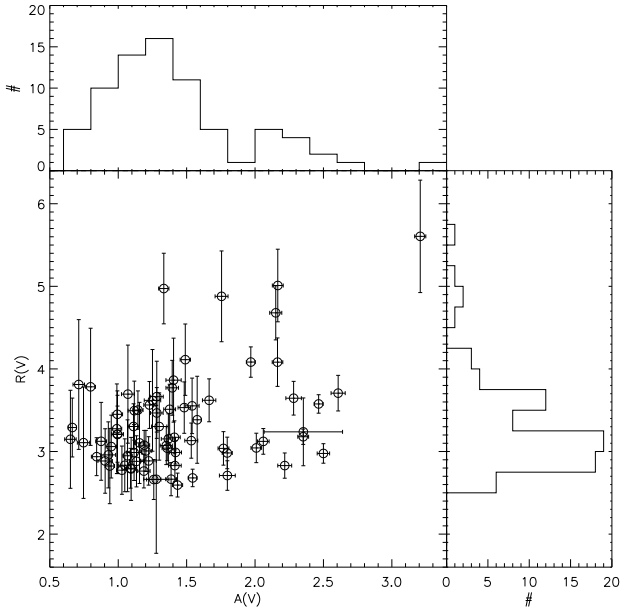}
\caption{The A(V) and R(V) values for each extinction curve are
plotted.  The histograms of both quantities also help
characterize the sample.
\label{fig_samp_char}}
\end{figure}

The FUSE+IUE extinction curves for all the stars in our sample are
shown in Fig.~\ref{fig_ext_curves1}.  The comparison stars used along
with details of each extinction curve are given in
Table~\ref{tab_ext_info}.  The quantities are given as value $\pm$
random $\pm$ systematic uncertainties.  The comparison star used for
each reddened star was picked to be the closest in the 2-dimensional
space defined by the MK spectral type.  We rejected any extinction
curves that had a reddened star to comparison spectral-type mismatch
larger than one temperature or luminosity subclass to account for
spectral type uncertainties.  The A(V) and R(V) distribution of our
sample are shown in Fig.~\ref{fig_samp_char}.

\section{Results}

\subsection{Comparison to \citet{Sofia05}}

The preliminary work on this topic using FUSE observations was
presented in \citet{Sofia05}.  The new extinction curves presented in
this paper improve on this previous work in a number of ways.
First, the calibration pipeline version used was 
v3.0 (compared to v2.2.1) which significantly improved the reduction
of faint sources, especially at the shorter FUSE wavelengths.  Second,
only sightlines where the $H_2$ modeling included populations beyond
J=1 were included.  Third, the set of comparison stars included more
spectral and luminosity types and the comparison star spectra were
carefully dereddened.  Finally, our new work imposed more stringent
data quality cuts and a requirement for close spectral type matches
between reddened and comparison stars.  These changes resulted in
three of the \citet{Sofia05} sightlines being rejected from our sample.
These three stars were HD~167971 (poor $H_2$ model), HD~210121 (no
good comparison star), and HD~239682 (poor FUSE and IUE data).

\begin{figure}[tbp]
\epsscale{1.1}
\plotone{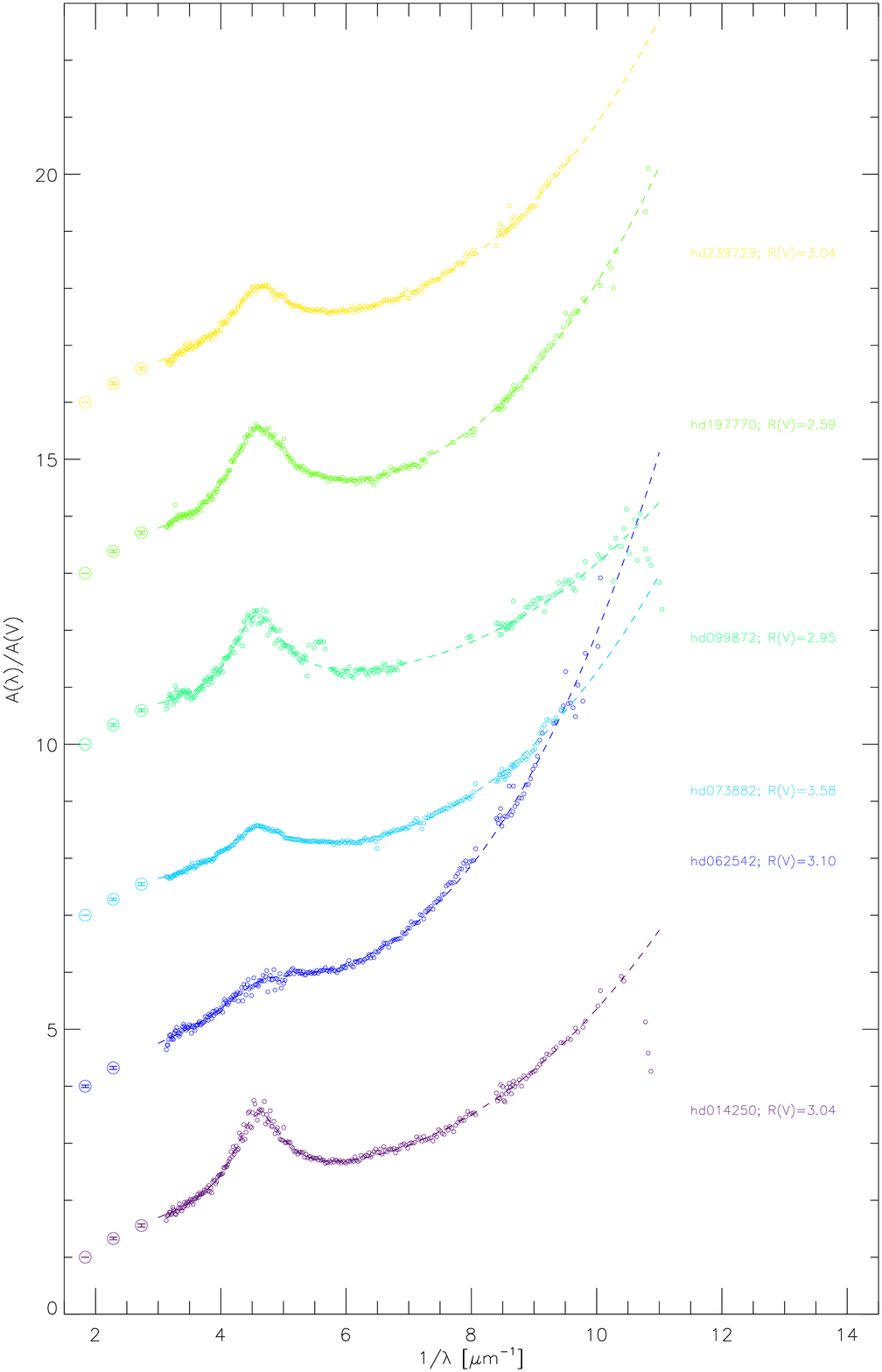}
\caption{The updated versions of 6 of the extinction curves presented in
\citet{Sofia05} are shown.
\label{fig_sofia_ext}}
\end{figure}

\begin{figure*}[tbp]
\epsscale{1.1}
\plotone{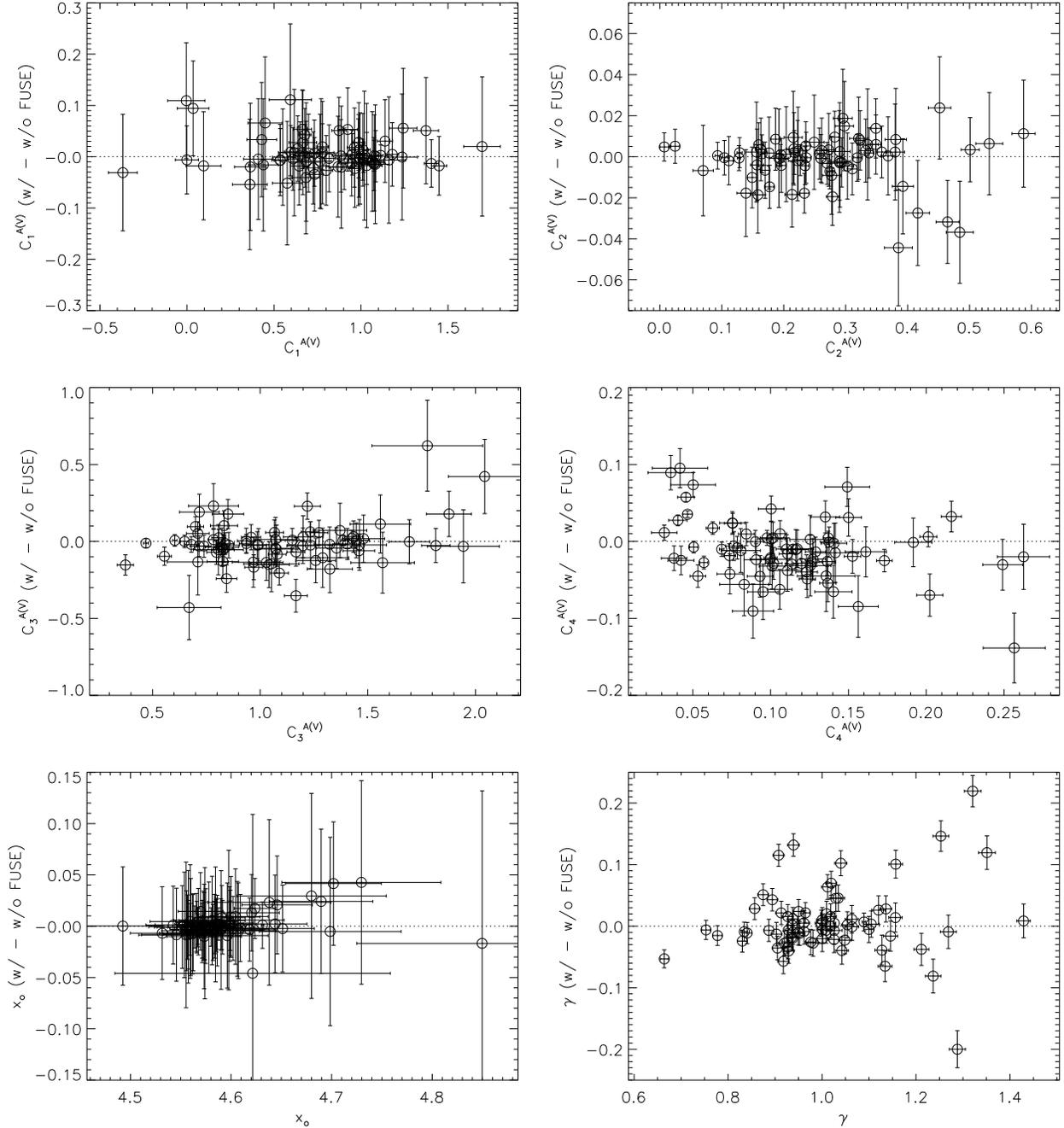}
\caption{The FM90 coefficients fit to the IUE+FUSE data are plotted
versus the difference between these coefficients fit with and without
the FUSE data.  In determining the fitting uncertainties for this
comparison, only the random uncertainties were considered.  The
systematic uncertainties affect the IUE+FUSE and IUE only curves in the
same manner. 
\label{fig_fm90_comp}}
\end{figure*}

The six stars from \citet{Sofia05} remaining in our sample are shown
in Fig.~\ref{fig_sofia_ext}.  These extinction curves are similar to
those shown in Fig.~2 of \citet{Sofia05}.  They extend to shorter
wavelengths and are generally less noisy.

\subsection{FM Fits}

\tabletypesize{\scriptsize}
\begin{deluxetable*}{rrrrrrr}
\tablewidth{0pt}
\tablecaption{FM90 Fit Parameters \label{tab_fm90_param}}
\tablehead{ \colhead{name} & \colhead{$C_1^{A(V)}$\tablenotemark{a}} & \colhead{$C_2^{A(V)}$\tablenotemark{a}} &
   \colhead{$C_3^{A(V)}$\tablenotemark{a}} & \colhead{$C_4^{A(V)}$\tablenotemark{a}} & \colhead{$x_o$\tablenotemark{a}} & \colhead{$\gamma$\tablenotemark{a}} }
\startdata
BD+35$\arcdeg$4258 & $ 0.60 \pm 0.119 \pm 0.032$ & $ 0.38 \pm 0.022 \pm 0.018$ & $ 1.84 \pm 0.276 \pm 0.078$ & $ 0.04 \pm 0.012 \pm 0.003$ & $ 4.73 \pm 0.080 \pm 0.012$ & $ 1.34 \pm 0.033 \pm 0.024$ \\
BD+53$\arcdeg$2820 & $ 0.70 \pm 0.060 \pm 0.026$ & $ 0.30 \pm 0.012 \pm 0.012$ & $ 1.06 \pm 0.089 \pm 0.042$ & $ 0.10 \pm 0.008 \pm 0.004$ & $ 4.62 \pm 0.033 \pm 0.012$ & $ 1.01 \pm 0.023 \pm 0.015$ \\
 BD+56$\arcdeg$524 & $ 0.67 \pm 0.046 \pm 0.018$ & $ 0.32 \pm 0.008 \pm 0.008$ & $ 1.40 \pm 0.078 \pm 0.036$ & $ 0.13 \pm 0.009 \pm 0.003$ & $ 4.58 \pm 0.021 \pm 0.008$ & $ 1.03 \pm 0.019 \pm 0.009$ \\
  HD001383 & $ 0.82 \pm 0.051 \pm 0.026$ & $ 0.26 \pm 0.011 \pm 0.008$ & $ 1.30 \pm 0.075 \pm 0.043$ & $ 0.11 \pm 0.008 \pm 0.004$ & $ 4.56 \pm 0.026 \pm 0.009$ & $ 1.00 \pm 0.019 \pm 0.010$ \\
  HD013268 & $ 0.53 \pm 0.057 \pm 0.019$ & $ 0.36 \pm 0.011 \pm 0.014$ & $ 0.83 \pm 0.062 \pm 0.039$ & $ 0.10 \pm 0.008 \pm 0.004$ & $ 4.58 \pm 0.028 \pm 0.016$ & $ 0.89 \pm 0.023 \pm 0.018$ \\
  HD014250 & $ 0.73 \pm 0.043 \pm 0.020$ & $ 0.29 \pm 0.008 \pm 0.008$ & $ 1.42 \pm 0.066 \pm 0.038$ & $ 0.13 \pm 0.007 \pm 0.003$ & $ 4.59 \pm 0.020 \pm 0.008$ & $ 1.00 \pm 0.018 \pm 0.009$ \\
  HD014434 & $ 0.41 \pm 0.082 \pm 0.023$ & $ 0.38 \pm 0.016 \pm 0.020$ & $ 1.24 \pm 0.118 \pm 0.070$ & $ 0.11 \pm 0.014 \pm 0.005$ & $ 4.56 \pm 0.035 \pm 0.008$ & $ 0.96 \pm 0.019 \pm 0.010$ \\
  HD015558 & $ 0.95 \pm 0.052 \pm 0.023$ & $ 0.23 \pm 0.010 \pm 0.007$ & $ 1.46 \pm 0.077 \pm 0.040$ & $ 0.13 \pm 0.009 \pm 0.003$ & $ 4.56 \pm 0.017 \pm 0.005$ & $ 1.00 \pm 0.017 \pm 0.005$ \\
  HD017505 & $ 0.61 \pm 0.045 \pm 0.022$ & $ 0.34 \pm 0.009 \pm 0.013$ & $ 1.40 \pm 0.069 \pm 0.053$ & $ 0.12 \pm 0.008 \pm 0.004$ & $ 4.57 \pm 0.020 \pm 0.006$ & $ 1.00 \pm 0.017 \pm 0.007$ \\
  HD023060 & $ 1.70 \pm 0.090 \pm 0.061$ & $ 0.07 \pm 0.015 \pm 0.005$ & $ 1.37 \pm 0.133 \pm 0.061$ & $ 0.14 \pm 0.010 \pm 0.006$ & $ 4.56 \pm 0.043 \pm 0.007$ & $ 1.12 \pm 0.026 \pm 0.019$ \\
  HD027778 & $ 0.87 \pm 0.071 \pm 0.028$ & $ 0.32 \pm 0.015 \pm 0.011$ & $ 1.31 \pm 0.102 \pm 0.051$ & $ 0.24 \pm 0.013 \pm 0.008$ & $ 4.59 \pm 0.045 \pm 0.008$ & $ 1.21 \pm 0.026 \pm 0.017$ \\
  HD037903 & $ 1.17 \pm 0.056 \pm 0.033$ & $ 0.11 \pm 0.011 \pm 0.007$ & $ 1.04 \pm 0.091 \pm 0.034$ & $ 0.14 \pm 0.009 \pm 0.004$ & $ 4.60 \pm 0.039 \pm 0.014$ & $ 1.19 \pm 0.026 \pm 0.018$ \\
  HD038087 & $ 1.39 \pm 0.029 \pm 0.042$ & $ 0.01 \pm 0.003 \pm 0.007$ & $ 0.88 \pm 0.082 \pm 0.031$ & $ 0.10 \pm 0.008 \pm 0.004$ & $ 4.56 \pm 0.034 \pm 0.009$ & $ 1.00 \pm 0.020 \pm 0.011$ \\
  HD045314 & $ 1.07 \pm 0.026 \pm 0.024$ & $ 0.13 \pm 0.006 \pm 0.006$ & $ 0.60 \pm 0.034 \pm 0.020$ & $ 0.06 \pm 0.005 \pm 0.002$ & $ 4.63 \pm 0.031 \pm 0.020$ & $ 1.02 \pm 0.028 \pm 0.023$ \\
  HD046056 & $ 0.73 \pm 0.054 \pm 0.025$ & $ 0.32 \pm 0.010 \pm 0.012$ & $ 1.17 \pm 0.065 \pm 0.048$ & $ 0.13 \pm 0.007 \pm 0.005$ & $ 4.57 \pm 0.025 \pm 0.014$ & $ 0.93 \pm 0.022 \pm 0.016$ \\
  HD046150 & $ 0.68 \pm 0.090 \pm 0.034$ & $ 0.30 \pm 0.018 \pm 0.016$ & $ 1.24 \pm 0.100 \pm 0.068$ & $ 0.15 \pm 0.014 \pm 0.007$ & $ 4.56 \pm 0.033 \pm 0.009$ & $ 0.92 \pm 0.017 \pm 0.008$ \\
  HD046202 & $ 0.80 \pm 0.055 \pm 0.026$ & $ 0.28 \pm 0.010 \pm 0.010$ & $ 1.07 \pm 0.057 \pm 0.044$ & $ 0.12 \pm 0.007 \pm 0.004$ & $ 4.58 \pm 0.029 \pm 0.015$ & $ 0.93 \pm 0.022 \pm 0.016$ \\
  HD047129 & $ 0.58 \pm 0.059 \pm 0.021$ & $ 0.27 \pm 0.012 \pm 0.011$ & $ 0.79 \pm 0.066 \pm 0.042$ & $ 0.07 \pm 0.008 \pm 0.003$ & $ 4.58 \pm 0.036 \pm 0.019$ & $ 0.93 \pm 0.028 \pm 0.023$ \\
  HD047240 & $ 0.76 \pm 0.084 \pm 0.029$ & $ 0.29 \pm 0.016 \pm 0.011$ & $ 1.25 \pm 0.127 \pm 0.047$ & $ 0.13 \pm 0.013 \pm 0.005$ & $ 4.64 \pm 0.046 \pm 0.012$ & $ 1.09 \pm 0.022 \pm 0.013$ \\
  HD047417 & $ 0.69 \pm 0.091 \pm 0.030$ & $ 0.29 \pm 0.017 \pm 0.013$ & $ 1.05 \pm 0.140 \pm 0.050$ & $ 0.13 \pm 0.015 \pm 0.006$ & $ 4.59 \pm 0.059 \pm 0.014$ & $ 1.04 \pm 0.023 \pm 0.016$ \\
  HD062542 & $ 0.38 \pm 0.086 \pm 0.013$ & $ 0.44 \pm 0.017 \pm 0.015$ & $ 0.79 \pm 0.126 \pm 0.038$ & $ 0.32 \pm 0.020 \pm 0.010$ & $ 4.64 \pm 0.079 \pm 0.018$ & $ 1.37 \pm 0.037 \pm 0.029$ \\
  HD073882 & $ 1.01 \pm 0.030 \pm 0.015$ & $ 0.20 \pm 0.007 \pm 0.004$ & $ 0.86 \pm 0.043 \pm 0.019$ & $ 0.18 \pm 0.008 \pm 0.003$ & $ 4.56 \pm 0.028 \pm 0.015$ & $ 1.16 \pm 0.025 \pm 0.017$ \\
  HD091651 & $ 0.68 \pm 0.078 \pm 0.031$ & $ 0.28 \pm 0.014 \pm 0.014$ & $ 0.84 \pm 0.095 \pm 0.054$ & $ 0.04 \pm 0.007 \pm 0.003$ & $ 4.65 \pm 0.058 \pm 0.026$ & $ 1.04 \pm 0.034 \pm 0.029$ \\
  HD093222 & $ 1.01 \pm 0.038 \pm 0.028$ & $ 0.13 \pm 0.008 \pm 0.006$ & $ 0.48 \pm 0.033 \pm 0.027$ & $ 0.06 \pm 0.005 \pm 0.002$ & $ 4.57 \pm 0.032 \pm 0.019$ & $ 0.85 \pm 0.026 \pm 0.022$ \\
  HD093250 & $ 0.89 \pm 0.096 \pm 0.031$ & $ 0.22 \pm 0.017 \pm 0.010$ & $ 0.98 \pm 0.126 \pm 0.043$ & $ 0.08 \pm 0.011 \pm 0.003$ & $ 4.55 \pm 0.060 \pm 0.011$ & $ 1.14 \pm 0.024 \pm 0.015$ \\
  HD093827 & $ 1.03 \pm 0.104 \pm 0.047$ & $ 0.19 \pm 0.019 \pm 0.010$ & $ 0.80 \pm 0.105 \pm 0.043$ & $ 0.10 \pm 0.013 \pm 0.005$ & $ 4.60 \pm 0.052 \pm 0.014$ & $ 0.92 \pm 0.021 \pm 0.014$ \\
  HD096675 & $ 1.24 \pm 0.105 \pm 0.048$ & $ 0.20 \pm 0.019 \pm 0.011$ & $ 2.00 \pm 0.184 \pm 0.084$ & $ 0.28 \pm 0.018 \pm 0.012$ & $ 4.56 \pm 0.043 \pm 0.010$ & $ 1.28 \pm 0.027 \pm 0.016$ \\
  HD096715 & $ 0.57 \pm 0.097 \pm 0.031$ & $ 0.30 \pm 0.018 \pm 0.016$ & $ 1.03 \pm 0.104 \pm 0.061$ & $ 0.16 \pm 0.015 \pm 0.008$ & $ 4.57 \pm 0.041 \pm 0.009$ & $ 0.91 \pm 0.017 \pm 0.008$ \\
  HD099872 & $ 1.15 \pm 0.095 \pm 0.037$ & $ 0.16 \pm 0.018 \pm 0.007$ & $ 1.42 \pm 0.131 \pm 0.053$ & $ 0.11 \pm 0.012 \pm 0.004$ & $ 4.57 \pm 0.037 \pm 0.005$ & $ 1.01 \pm 0.021 \pm 0.012$ \\
  HD099890 & $ 0.00 \pm 0.000 \pm 0.025$ & $ 0.49 \pm 0.009 \pm 0.031$ & $ 0.60 \pm 0.106 \pm 0.055$ & $ 0.02 \pm 0.006 \pm 0.003$ & $ 4.66 \pm 0.077 \pm 0.016$ & $ 0.82 \pm 0.021 \pm 0.017$ \\
  HD100213 & $ 1.08 \pm 0.077 \pm 0.040$ & $ 0.16 \pm 0.014 \pm 0.009$ & $ 0.97 \pm 0.082 \pm 0.043$ & $ 0.11 \pm 0.008 \pm 0.005$ & $ 4.60 \pm 0.034 \pm 0.018$ & $ 0.95 \pm 0.024 \pm 0.018$ \\
  HD101190 & $ 0.44 \pm 0.091 \pm 0.029$ & $ 0.34 \pm 0.018 \pm 0.022$ & $ 1.06 \pm 0.114 \pm 0.075$ & $ 0.13 \pm 0.016 \pm 0.008$ & $ 4.59 \pm 0.046 \pm 0.011$ & $ 0.93 \pm 0.019 \pm 0.011$ \\
  HD101205 & $ 0.96 \pm 0.066 \pm 0.036$ & $ 0.19 \pm 0.013 \pm 0.009$ & $ 0.79 \pm 0.065 \pm 0.043$ & $ 0.10 \pm 0.009 \pm 0.004$ & $ 4.56 \pm 0.037 \pm 0.019$ & $ 0.90 \pm 0.028 \pm 0.023$ \\
  HD103779 & $ 0.99 \pm 0.061 \pm 0.051$ & $ 0.22 \pm 0.012 \pm 0.014$ & $ 0.86 \pm 0.062 \pm 0.049$ & $ 0.14 \pm 0.010 \pm 0.008$ & $ 4.56 \pm 0.030 \pm 0.013$ & $ 0.84 \pm 0.018 \pm 0.011$ \\
  HD122879 & $ 1.15 \pm 0.082 \pm 0.036$ & $ 0.15 \pm 0.015 \pm 0.007$ & $ 0.53 \pm 0.058 \pm 0.028$ & $ 0.09 \pm 0.011 \pm 0.003$ & $ 4.57 \pm 0.038 \pm 0.010$ & $ 0.74 \pm 0.015 \pm 0.009$ \\
  HD124979 & $ 0.64 \pm 0.076 \pm 0.026$ & $ 0.35 \pm 0.014 \pm 0.015$ & $ 0.71 \pm 0.060 \pm 0.039$ & $ 0.12 \pm 0.011 \pm 0.005$ & $ 4.57 \pm 0.034 \pm 0.017$ & $ 0.77 \pm 0.023 \pm 0.019$ \\
  HD147888 & $ 1.43 \pm 0.045 \pm 0.024$ & $ 0.03 \pm 0.006 \pm 0.003$ & $ 0.74 \pm 0.072 \pm 0.026$ & $ 0.07 \pm 0.007 \pm 0.003$ & $ 4.58 \pm 0.033 \pm 0.007$ & $ 0.94 \pm 0.023 \pm 0.017$ \\
  HD148422 & $ 0.01 \pm 0.003 \pm 0.015$ & $ 0.50 \pm 0.006 \pm 0.024$ & $ 0.84 \pm 0.076 \pm 0.049$ & $ 0.11 \pm 0.010 \pm 0.005$ & $ 4.61 \pm 0.039 \pm 0.014$ & $ 0.93 \pm 0.020 \pm 0.012$ \\
  HD149404 & $ 1.08 \pm 0.036 \pm 0.028$ & $ 0.16 \pm 0.007 \pm 0.005$ & $ 1.09 \pm 0.056 \pm 0.029$ & $ 0.14 \pm 0.007 \pm 0.004$ & $ 4.57 \pm 0.024 \pm 0.007$ & $ 1.03 \pm 0.018 \pm 0.007$ \\
  HD151805 & $ 1.00 \pm 0.079 \pm 0.038$ & $ 0.17 \pm 0.015 \pm 0.008$ & $ 1.00 \pm 0.095 \pm 0.044$ & $ 0.12 \pm 0.010 \pm 0.005$ & $ 4.55 \pm 0.037 \pm 0.011$ & $ 0.94 \pm 0.019 \pm 0.011$ \\
  HD152233 & $ 0.61 \pm 0.098 \pm 0.029$ & $ 0.29 \pm 0.020 \pm 0.015$ & $ 1.06 \pm 0.117 \pm 0.058$ & $ 0.09 \pm 0.013 \pm 0.004$ & $ 4.58 \pm 0.047 \pm 0.010$ & $ 0.99 \pm 0.020 \pm 0.010$ \\
  HD152234 & $ 0.77 \pm 0.074 \pm 0.028$ & $ 0.26 \pm 0.013 \pm 0.011$ & $ 0.83 \pm 0.081 \pm 0.034$ & $ 0.10 \pm 0.010 \pm 0.004$ & $ 4.55 \pm 0.036 \pm 0.010$ & $ 0.91 \pm 0.018 \pm 0.009$ \\
  HD152248 & $ 0.91 \pm 0.063 \pm 0.027$ & $ 0.22 \pm 0.012 \pm 0.008$ & $ 0.70 \pm 0.055 \pm 0.032$ & $ 0.08 \pm 0.009 \pm 0.003$ & $ 4.59 \pm 0.033 \pm 0.016$ & $ 0.85 \pm 0.023 \pm 0.018$ \\
  HD152249 & $ 1.00 \pm 0.066 \pm 0.022$ & $ 0.20 \pm 0.011 \pm 0.005$ & $ 0.99 \pm 0.071 \pm 0.023$ & $ 0.10 \pm 0.007 \pm 0.002$ & $ 4.60 \pm 0.032 \pm 0.009$ & $ 1.02 \pm 0.020 \pm 0.010$ \\
  HD152723 & $ 0.65 \pm 0.057 \pm 0.028$ & $ 0.24 \pm 0.011 \pm 0.012$ & $ 0.92 \pm 0.065 \pm 0.051$ & $ 0.08 \pm 0.009 \pm 0.004$ & $ 4.57 \pm 0.031 \pm 0.009$ & $ 0.93 \pm 0.018 \pm 0.010$ \\
  HD157857 & $ 0.78 \pm 0.088 \pm 0.032$ & $ 0.22 \pm 0.017 \pm 0.011$ & $ 1.47 \pm 0.118 \pm 0.068$ & $ 0.11 \pm 0.012 \pm 0.005$ & $ 4.54 \pm 0.029 \pm 0.007$ & $ 0.96 \pm 0.018 \pm 0.009$ \\
  HD160993 & $ 0.10 \pm 0.028 \pm 0.019$ & $ 0.53 \pm 0.011 \pm 0.035$ & $ 0.80 \pm 0.133 \pm 0.064$ & $ 0.16 \pm 0.018 \pm 0.011$ & $ 4.70 \pm 0.079 \pm 0.018$ & $ 0.98 \pm 0.023 \pm 0.017$ \\
  HD163522 & $-0.36 \pm 0.095 \pm 0.028$ & $ 0.59 \pm 0.022 \pm 0.035$ & $ 0.70 \pm 0.145 \pm 0.054$ & $ 0.10 \pm 0.018 \pm 0.007$ & $ 4.85 \pm 0.084 \pm 0.023$ & $ 1.22 \pm 0.033 \pm 0.026$ \\
  HD164816 & $ 1.24 \pm 0.095 \pm 0.056$ & $ 0.14 \pm 0.017 \pm 0.011$ & $ 0.84 \pm 0.086 \pm 0.044$ & $ 0.06 \pm 0.010 \pm 0.004$ & $ 4.60 \pm 0.042 \pm 0.012$ & $ 0.88 \pm 0.019 \pm 0.012$ \\
  HD164906 & $ 0.98 \pm 0.039 \pm 0.021$ & $ 0.10 \pm 0.007 \pm 0.005$ & $ 0.62 \pm 0.037 \pm 0.021$ & $ 0.05 \pm 0.004 \pm 0.002$ & $ 4.58 \pm 0.029 \pm 0.012$ & $ 1.00 \pm 0.021 \pm 0.013$ \\
  HD165052 & $ 0.96 \pm 0.074 \pm 0.048$ & $ 0.17 \pm 0.013 \pm 0.011$ & $ 1.04 \pm 0.079 \pm 0.066$ & $ 0.08 \pm 0.008 \pm 0.004$ & $ 4.57 \pm 0.028 \pm 0.009$ & $ 0.94 \pm 0.020 \pm 0.012$ \\
  HD167402 & $ 0.02 \pm 0.006 \pm 0.022$ & $ 0.47 \pm 0.009 \pm 0.023$ & $ 0.73 \pm 0.111 \pm 0.045$ & $ 0.04 \pm 0.010 \pm 0.003$ & $ 4.64 \pm 0.066 \pm 0.014$ & $ 0.91 \pm 0.022 \pm 0.015$ \\
  HD167771 & $ 1.03 \pm 0.061 \pm 0.032$ & $ 0.17 \pm 0.012 \pm 0.008$ & $ 0.93 \pm 0.063 \pm 0.043$ & $ 0.10 \pm 0.008 \pm 0.004$ & $ 4.56 \pm 0.030 \pm 0.016$ & $ 0.92 \pm 0.026 \pm 0.021$ \\
  HD168076 & $ 0.69 \pm 0.058 \pm 0.015$ & $ 0.23 \pm 0.011 \pm 0.006$ & $ 1.25 \pm 0.104 \pm 0.035$ & $ 0.05 \pm 0.008 \pm 0.002$ & $ 4.63 \pm 0.033 \pm 0.007$ & $ 1.16 \pm 0.022 \pm 0.010$ \\
  HD168941 & $ 0.99 \pm 0.093 \pm 0.034$ & $ 0.15 \pm 0.018 \pm 0.007$ & $ 0.41 \pm 0.060 \pm 0.025$ & $ 0.06 \pm 0.012 \pm 0.003$ & $ 4.53 \pm 0.053 \pm 0.012$ & $ 0.69 \pm 0.015 \pm 0.010$ \\
  HD178487 & $ 0.92 \pm 0.073 \pm 0.029$ & $ 0.21 \pm 0.014 \pm 0.008$ & $ 0.83 \pm 0.079 \pm 0.035$ & $ 0.10 \pm 0.011 \pm 0.004$ & $ 4.60 \pm 0.035 \pm 0.010$ & $ 0.88 \pm 0.018 \pm 0.010$ \\
  HD179406 & $ 1.03 \pm 0.101 \pm 0.039$ & $ 0.25 \pm 0.019 \pm 0.011$ & $ 1.57 \pm 0.142 \pm 0.064$ & $ 0.16 \pm 0.014 \pm 0.007$ & $ 4.61 \pm 0.036 \pm 0.006$ & $ 1.05 \pm 0.021 \pm 0.011$ \\
  HD179407 & $ 0.43 \pm 0.083 \pm 0.019$ & $ 0.39 \pm 0.016 \pm 0.017$ & $ 1.57 \pm 0.169 \pm 0.068$ & $ 0.13 \pm 0.015 \pm 0.005$ & $ 4.71 \pm 0.056 \pm 0.015$ & $ 1.26 \pm 0.029 \pm 0.020$ \\
  HD185418 & $ 1.37 \pm 0.094 \pm 0.042$ & $ 0.16 \pm 0.017 \pm 0.009$ & $ 1.89 \pm 0.118 \pm 0.060$ & $ 0.14 \pm 0.011 \pm 0.004$ & $ 4.59 \pm 0.025 \pm 0.007$ & $ 1.03 \pm 0.020 \pm 0.010$ \\
  HD188001 & $ 0.67 \pm 0.091 \pm 0.028$ & $ 0.28 \pm 0.017 \pm 0.013$ & $ 0.70 \pm 0.087 \pm 0.048$ & $ 0.04 \pm 0.010 \pm 0.003$ & $ 4.58 \pm 0.052 \pm 0.021$ & $ 0.85 \pm 0.030 \pm 0.026$ \\
  HD190603 & $ 0.54 \pm 0.062 \pm 0.066$ & $ 0.33 \pm 0.013 \pm 0.041$ & $ 1.35 \pm 0.117 \pm 0.166$ & $ 0.04 \pm 0.008 \pm 0.005$ & $ 4.65 \pm 0.038 \pm 0.007$ & $ 1.15 \pm 0.021 \pm 0.009$ \\
  HD192639 & $ 0.78 \pm 0.049 \pm 0.017$ & $ 0.27 \pm 0.010 \pm 0.007$ & $ 1.23 \pm 0.070 \pm 0.040$ & $ 0.06 \pm 0.008 \pm 0.002$ & $ 4.57 \pm 0.021 \pm 0.010$ & $ 0.96 \pm 0.023 \pm 0.016$ \\
  HD197770 & $ 0.99 \pm 0.045 \pm 0.027$ & $ 0.23 \pm 0.009 \pm 0.007$ & $ 1.81 \pm 0.074 \pm 0.053$ & $ 0.21 \pm 0.006 \pm 0.006$ & $ 4.59 \pm 0.015 \pm 0.008$ & $ 1.10 \pm 0.017 \pm 0.011$ \\
  HD198781 & $ 0.44 \pm 0.104 \pm 0.025$ & $ 0.42 \pm 0.020 \pm 0.018$ & $ 2.09 \pm 0.225 \pm 0.084$ & $ 0.13 \pm 0.017 \pm 0.006$ & $ 4.69 \pm 0.057 \pm 0.013$ & $ 1.36 \pm 0.031 \pm 0.021$ \\
  HD199579 & $ 0.37 \pm 0.085 \pm 0.025$ & $ 0.38 \pm 0.018 \pm 0.024$ & $ 1.28 \pm 0.141 \pm 0.088$ & $ 0.14 \pm 0.017 \pm 0.009$ & $ 4.57 \pm 0.052 \pm 0.013$ & $ 1.12 \pm 0.024 \pm 0.015$ \\
  HD200775 & $ 1.08 \pm 0.036 \pm 0.014$ & $ 0.11 \pm 0.006 \pm 0.002$ & $ 0.75 \pm 0.058 \pm 0.028$ & $ 0.09 \pm 0.005 \pm 0.002$ & $ 4.55 \pm 0.043 \pm 0.007$ & $ 1.40 \pm 0.034 \pm 0.026$ \\
  HD203938 & $ 0.95 \pm 0.025 \pm 0.017$ & $ 0.23 \pm 0.005 \pm 0.006$ & $ 1.39 \pm 0.046 \pm 0.030$ & $ 0.14 \pm 0.005 \pm 0.003$ & $ 4.57 \pm 0.011 \pm 0.006$ & $ 1.07 \pm 0.015 \pm 0.008$ \\
  HD206267 & $ 0.55 \pm 0.067 \pm 0.026$ & $ 0.37 \pm 0.013 \pm 0.017$ & $ 1.68 \pm 0.101 \pm 0.081$ & $ 0.20 \pm 0.013 \pm 0.009$ & $ 4.58 \pm 0.026 \pm 0.008$ & $ 1.06 \pm 0.020 \pm 0.010$ \\
  HD206773 & $ 0.84 \pm 0.070 \pm 0.020$ & $ 0.18 \pm 0.012 \pm 0.007$ & $ 0.75 \pm 0.073 \pm 0.025$ & $ 0.04 \pm 0.007 \pm 0.002$ & $ 4.57 \pm 0.041 \pm 0.010$ & $ 1.04 \pm 0.023 \pm 0.016$ \\
  HD207198 & $ 0.73 \pm 0.049 \pm 0.019$ & $ 0.35 \pm 0.010 \pm 0.008$ & $ 1.16 \pm 0.068 \pm 0.028$ & $ 0.20 \pm 0.010 \pm 0.005$ & $ 4.62 \pm 0.027 \pm 0.009$ & $ 1.04 \pm 0.018 \pm 0.008$ \\
  HD209339 & $ 1.08 \pm 0.089 \pm 0.047$ & $ 0.22 \pm 0.017 \pm 0.011$ & $ 1.03 \pm 0.081 \pm 0.048$ & $ 0.11 \pm 0.010 \pm 0.005$ & $ 4.57 \pm 0.034 \pm 0.012$ & $ 0.91 \pm 0.019 \pm 0.011$ \\
  HD216898 & $ 1.11 \pm 0.014 \pm 0.021$ & $ 0.20 \pm 0.003 \pm 0.005$ & $ 1.36 \pm 0.041 \pm 0.036$ & $ 0.10 \pm 0.005 \pm 0.002$ & $ 4.59 \pm 0.012 \pm 0.009$ & $ 0.99 \pm 0.019 \pm 0.012$ \\
  HD239729 & $ 0.87 \pm 0.049 \pm 0.021$ & $ 0.26 \pm 0.009 \pm 0.008$ & $ 1.16 \pm 0.060 \pm 0.029$ & $ 0.18 \pm 0.008 \pm 0.004$ & $ 4.60 \pm 0.022 \pm 0.009$ & $ 1.11 \pm 0.021 \pm 0.011$ \\
  HD326329 & $ 1.07 \pm 0.074 \pm 0.077$ & $ 0.27 \pm 0.013 \pm 0.020$ & $ 1.20 \pm 0.089 \pm 0.093$ & $ 0.12 \pm 0.009 \pm 0.009$ & $ 4.61 \pm 0.033 \pm 0.016$ & $ 0.95 \pm 0.024 \pm 0.018$ \\
  HD332407 & $ 0.64 \pm 0.050 \pm 0.025$ & $ 0.31 \pm 0.009 \pm 0.011$ & $ 1.27 \pm 0.067 \pm 0.051$ & $ 0.10 \pm 0.007 \pm 0.004$ & $ 4.57 \pm 0.024 \pm 0.010$ & $ 1.01 \pm 0.019 \pm 0.011$
\enddata
\tablenotetext{a}{The quantitities are given as value $\pm$ random uncertainty $\pm$ systematic uncertainty.}
\end{deluxetable*}

Each $A(\lambda)/A(V)$ curve was fit with the FM90 parameterization of
the UV extinction curve \citep{Fitzpatrick86, Fitzpatrick88,
Fitzpatrick90}.  This parameterization 
was developed to describe the extinction curve in the IUE spectral
range (1150 -- 3300~\AA).  The FM90 function, modified to fit
$A(\lambda)/A(V)$ curves, is:
\begin{equation}
\frac{A(\lambda)}{A(V)} =
   C_1^{A(V)} + C_2^{A(V)}x + C_3^{A(V)}D(x, \gamma, x_o) + C_4^{A(V)}F(x)
\end{equation}
where $x = 1/\lambda [\micron^{-1}]$, the 2175~\AA\ bump is
represented by the Drude term 
\begin{equation}
D(x, \gamma, x_o) = \frac{x^2}{(x^2 - x_o^2)^2 + (x\gamma)^2},
\end{equation}
and the far-UV curvature (for $x \ge 5.9~\micron^{-1}$) is given by
\begin{equation}
F(x) = 0.5392(x - 5.9)^2 + 0.05644(x - 5.9)^3
\end{equation}
These 6 parameters are not uncorrelated and
determining the best fit requires iteration to converge to a stable
solution.  
We have found that a fitting algorithm that alternates
between fitting the four coefficients ($C_1^{A(V)}, C_2^{A(V)},
C_3^{A(V)}, \& C_4^{A(V)}$), 
$\gamma$, and $x_o$ produces the most stable and best quality fits.
As part of this fitting algorithm, we also have found that iterative
sigma clipping is important to ensure that a small number of highly
deviant points does not bias the fits.  The relationship between our
version of the FM90 parameters and those fit to $E(\lambda-V)/E(B-V)$
\citep{Fitzpatrick90} is 
\begin{eqnarray}
C_1^{A(V)} & = & C_1/R(V) + 1 \\
C_j^{A(V)} & = & C_j/R(V),~\mathrm{for}~j = 2,3,4.
\end{eqnarray}
The uncertainties on the resulting FM90 parameters are composed of
random and systematic components.  The random component is due to the
uncertainties on the measurement of the ultraviolet fluxes and the
systematic component is due to uncertainties in the V band magnitudes
and in the calculation of $A(V)$.  The random uncertainties were
calculated using the Monte Carlo technique generating trials near the
best fit and determining when the resulting FM90 fit is within the
errors using the F-test method.  The random uncertainty is then
one half the difference between the minimum and maximum fit values that
satisfy the F-test criteria.  The systematic uncertainties are
determined by one half the difference of the FM90 fit coefficients
determined from the two curves generated by adding and subtracting the
systematic uncertainties from the curve.  Finally, the random and
systematic uncertainties can be combined in quadrature to produce the
final FM90 parameter uncertainties.

The FM90 fit coefficients for the extinction curves are given in
Table~\ref{tab_fm90_param} with the value $\pm$ random $\pm$
systematic uncertainties.

Fitting the IUE+FUSE extinction curve with the FM90 parameterization does
produce different FM90 fit coefficients than fitting just the IUE data.  The
effect of adding the FUSE data on the FM90 fit coefficients is shown
in Fig.~\ref{fig_fm90_comp}.  The FM90 coefficients that show
significant change are $C_4^{A(V)}$ and $\gamma$.  This is not a surprising
result as the $C_4^{A(V)}$ coefficient describes the strength of the far-UV
rise and $\gamma$ is the width of the 2175~\AA\ bump.  The FUSE data put
much stronger constraints on the far-UV rise and this, in turn,
affects the best fit width of the bump.  The most significant result
is that there is an average shift in the value of $C_4^{A(V)}$ of
$0.0072 \pm 0.0016$ which is a $\sim$8\% reduction in the average
strength on the far-UV rise.  None of the other parameters have
significant shifts in their average values.  Finally, the addition of
the FUSE data removes the negative far-UV curvature strengths
($C_4^{A(V)}$) that were seen with fits to only the IUE data.

\begin{figure}[tbp]
\epsscale{1.1}
\plotone{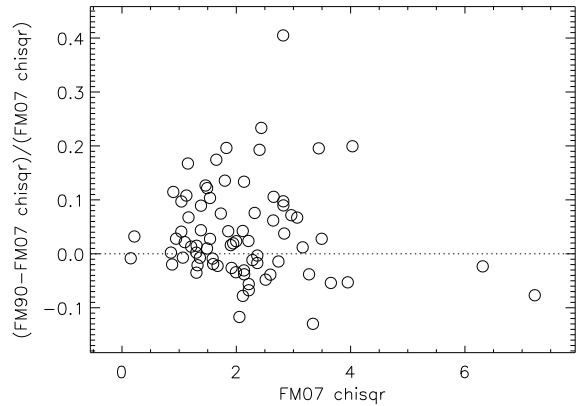}
\caption{The fractional difference between the $\chi^2$ values for the
FM90 and FM07 fits are shown.  On average, the FM07 $\chi^2$ is
better, with the split of which fit produces the smaller $\chi^2$
being 46 to 30 between FM07 and FM90, respectively. 
\label{fig_fm90_fm07_comp}}
\end{figure}

Recently, \citet{Fitzpatrick07} presented a modification to the FM90
parameterization (hereafter called FM07).  They introduced a 7th
coefficient ($C_5$) to allow the wavelength where the far-UV curvature
term is important to be a fitted parameter.  They also modified the
far-UV term to just be a quadratic (removing the cubic term).  We have
compared the fit $\chi^2$ values between FM90 and FM07 fits to the
IUE+FUSE extinction curves in Fig.~\ref{fig_fm90_fm07_comp}.  On
average, there is a preference for the FM07 over the FM90
parameterization, but this is a weak preference given that 40\% of the
sample is better fit with the FM90 parameterization.  There are
significant differences between some of the parameters ($C_3^{A(V)}$,
$C_4^{A(V)}$, and $\gamma$) between the FM90 and FM07 fits, but it
could be that this is just a reflection of the non-orthogonal nature
of the fit parameters and/or the result of adding an additional free
parameter.  Since there is no strong reason to prefer the newer FM07
parameterization over the FM90 parameterization, we use the FM90
parameterization for the majority of this paper as it has one fewer
free parameter and is easier to compare with previous work.

\subsection{$R(V)$ Dependent Relationship}
\label{sec_rv_rel}

\begin{figure*}[tbp]
\epsscale{1.1}
\plotone{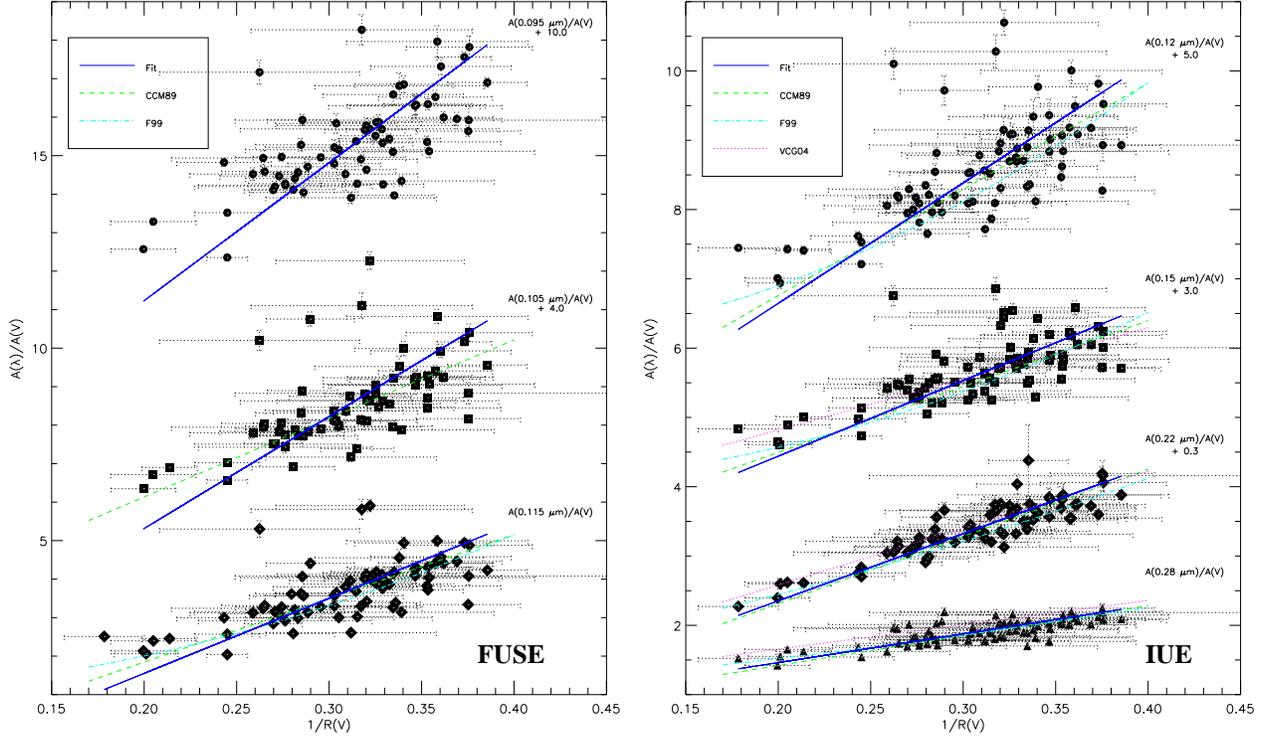}
\caption{The relationship between $A(\lambda)/A(V)$ and $R(V)^{-1}$ is
shown for selected wavelengths using FUSE (left) and IUE (right).  The
relationships given by the CCM89 ($\lambda < 10~\micron^{-1}$), F99
($\lambda < 8.7~\micron^{-1}$), and VCG04 ($\lambda <
8~\micron^{-1}$) fits are also shown,
but only for wavelengths for which they are valid.
The IUE plot is similar to that of Fig.~1a of CCM89.
The F99 relationship is non-linear in this form as it is based on
correlations between FM90 fit parameters and $R(V)^{-1}$.
\label{fig_fuse_ccm_alav_rv}}
\end{figure*}

\begin{figure*}[tbp]
\epsscale{1.1}
\plottwo{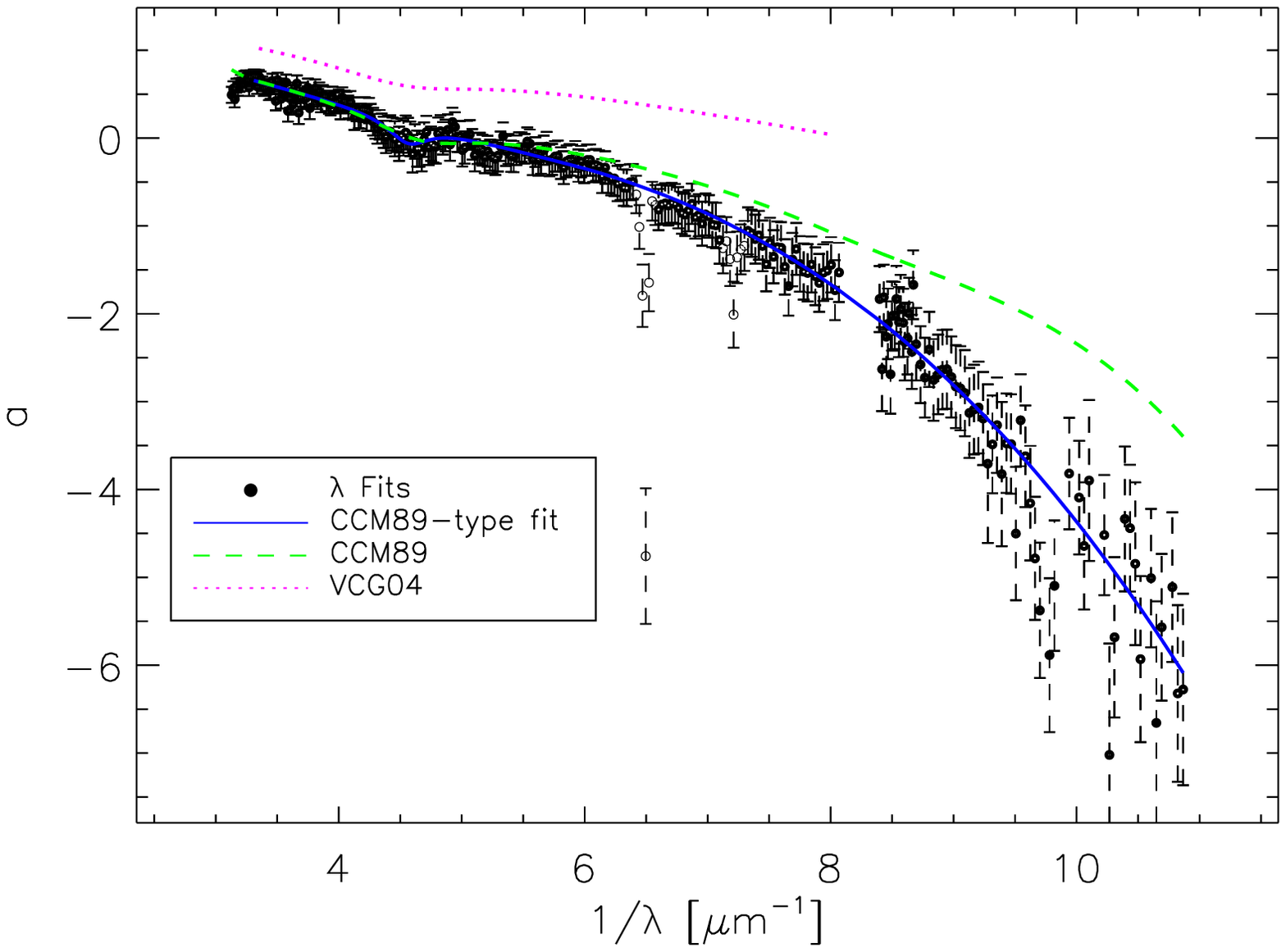}{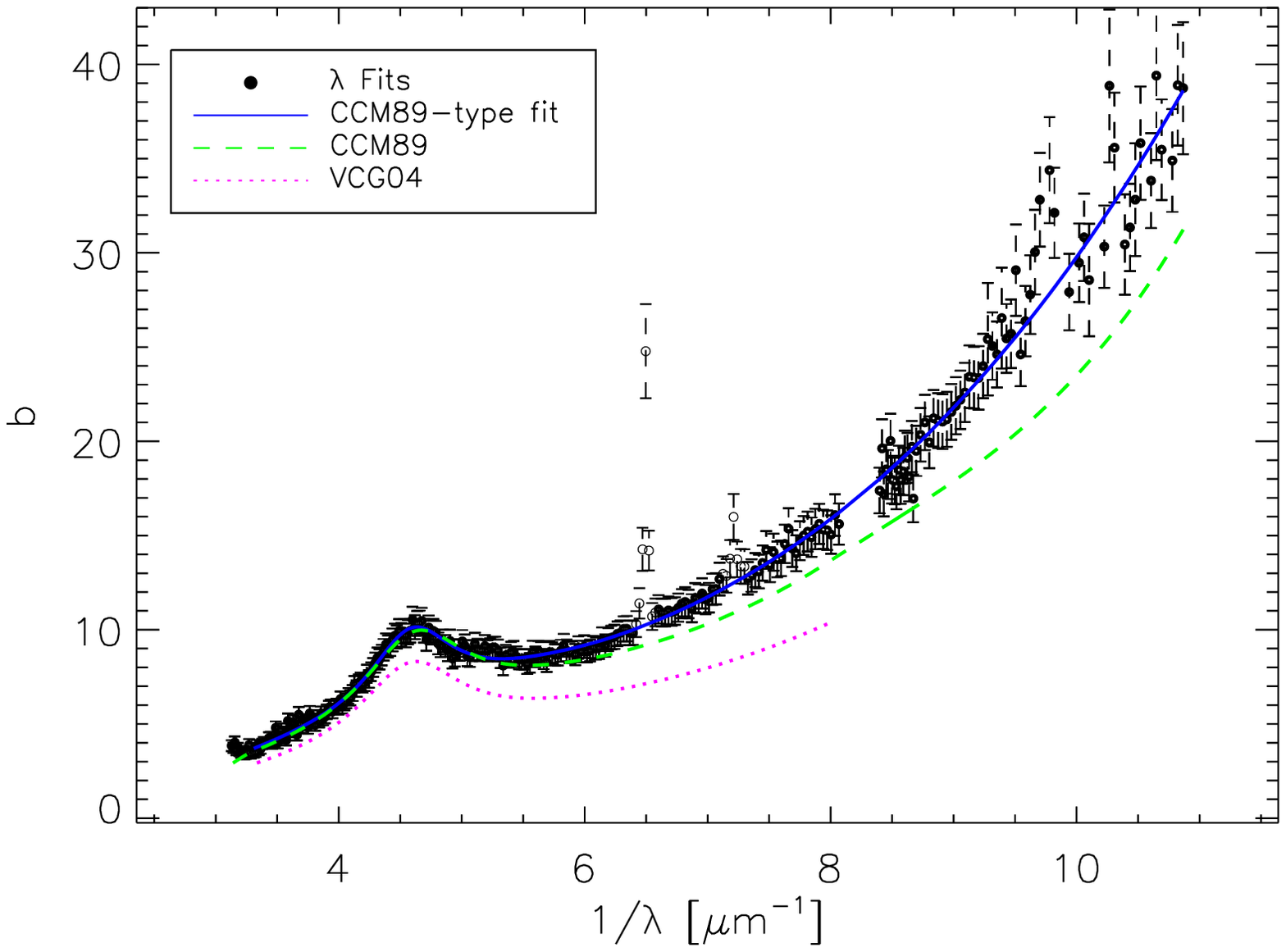}
\caption{The intercept (a) and slope (b) coefficients to the fits of
$A(\lambda)/A(V)$ versus $R(V)^{-1}$ are plotted versus wavelength.
Only fits using 40 or more points are plotted.
The coefficients for the individual fits ($\lambda$ fits) and the
CCM89-type functional fit to them are shown.  The coefficients from
CCM89 and VCG04 are also given for comparison.  The regions around the
wind lines (6.5 and 7.1 $\micron^{-1}$, open circles) are not used in the CCM89-like
fit as they are almost never well matched by the comparison stars.
\label{fig_fuse_ccm_a_b}}
\end{figure*}

The work of CCM89 demonstrated that {\it on average} extinction curves
follow a family of curves that can be described with one parameter.
Real deviations from the CCM89 relationship exist and were studied by
\citet{Mathis92} and VCG04.  CCM89 chose $R(V)$ as the one parameter
as it roughly measures the average grain size.  The CCM89 relationship
is an empirical relationship based on determining the linear
correlation between $A(\lambda)/A(V)$ and $R(V)^{-1}$ at each
wavelength with the resulting set of intercepts and slopes forming the
the CCM89 $R(V)$ dependent relationship.  The CCM89 relationship was
constructed mainly using \citet{Fitzpatrick88} fits to IUE extinction
curves and broadband data in the near-infrared and optical of 29
sightlines.  This was supplemented with ANS UV photometry and
Copernicus far-UV spectra for a smaller number of sightlines.  The
CCM89 relationship in the far-UV (8--10 $\micron^{-1}$) is mainly
based on extrapolating the \citet{Fitzpatrick88} fits and represents
the the most uncertain portion of the CCM89 relationship.  Overall,
CCM89 is valid from 0.3--10 $\micron^{-1}$ (3.3--0.1 $\micron$).

Since CCM89, there have been two other efforts to refine the R(V)
dependent relationship in the ultraviolet \citep{Fitzpatrick99review,
Valencic04}. \citet{Fitzpatrick99review} (hereafter F99) used the FM90
set of IUE extinction curves as the basis for their $R(V)$ dependent
relationship.  Their sample included 77 sightlines and the F99
relationship in the ultraviolet is based on the correlation between
$C_1$ and $C_2$ FM90 coefficients and $R(V)^{-1}$ (the other coefficients
are held constant).  The near-IR and optical regions are fit using
splines to account for the broad-band nature (and shifting effective
wavelengths) of the data.  The F99 relationship is valid from
$\sim$0.2--8.7 $\micron^{-1}$ (5.0--0.115 $\micron$).

VCG04 used a sample of 417 IUE
extinction curves as the basis for their $R(V)$ dependent
relationship.  They followed the methodology of CCM89 and fit
$A(\lambda)/A(V)$ and $R(V)^{-1}$ at each wavelength.  The resulting
slopes and intercepts were fit with the same functional form as CCM89.
This work concentrated on the IUE UV region and this is only valid
between 3.3--8.0 $\micron^{-1}$ (0.3--0.125 $\micron$).

Using our sample of 75 FUSE extinction curves we can investigate the
$R(V)$ dependent relationship in the far-UV (8.4--11 $\micron^{-1}$).
We start by following the CCM89 methodology and perform fits of the
data to
\begin{equation}
\frac{A(x)}{A(V)} = a(x) + \frac{b(x)}{R(V)}
\end{equation}
where $x = \lambda^{-1}$.  For
consistency, we have determined the linear fit coefficients for both
our FUSE and IUE extinction curves as this allows us to directly
compare our $R(V)$ dependent relationship to the CCM89, F99, and
VCG04.  Unlike CCM89, we 
perform our linear fits on the actual extinction curves, not the FM90
fits.  This directly connects our $R(V)$ relationship to the data
and does not assume a functional form for the UV extinction curve.
The linear fits are done using the 'fitexy' IDL program that takes
into account uncertainties on both the independent and dependent
variables.  Another approach was taken in deriving the $R(V)$ dependent
relationship by F99 who correlated the FM90 parameters with $R(V)$.
We did a similar analysis and found, just as F99 did, that most of the
dependence of the extinction curves on $R(V)$ is due to correlations with
$C_1$ and $C_2$.  We choose to use the CCM89 methodology for our
$R(V)$ relationship as it imposes fewer assumptions about the detailed
structure of the relationship.  Both approaches (CCM89 or F99) produce
very similar $R(V)$ relationships.

The fit relationships, fits, and comparisons to previous
$R(V)$ dependent relationships are shown in
Fig.~\ref{fig_fuse_ccm_alav_rv}.  The relationship shown by our
sample of sightlines is similar to those seen in previous works, but
our fits display a stronger dependence on $R(V)^{-1}$.  In
addition, our $R(V)$ dependent relationship is well defined all the way
to 11~$\micron^{-1}$.  As can be seen from this figure, our $R(V)$
dependent relationship is valid for $R(V)$ values between 2.5 and
$\sim$5.  The intercept (a) and slope (b) values are shown as a function
of wavelength in Fig.~\ref{fig_fuse_ccm_a_b} along with the values from
previous works.  We find quite different values for a and b compared
to previous works and this illustrates that the values of a and b are
not independent given that the differences between the different works
are not nearly as striking in Fig.~\ref{fig_fuse_ccm_alav_rv}.
As a check of the use of extrapolated FM90 parameters
to deredden the comparison stars (\S\ref{sec_comp_stars}), we have
verified that the linear fits at each wavelength do not change
significantly even when sightlines with lower reddenings are excluded
from the fit (up to $A(V) = 1.5$).

We fit our $a(x)$ and $b(x)$ values using a similar
functional form as given by CCM89.  Given that we have seen in this
work that the FM90 parameterization of the far-ultraviolet ($x >
5.9~\micron^{-1}$) provides a good representation of extinction curves
to $x = 11~\micron^{-1}$, we have expanded the term used by CCM89 to
represent $3.3~\micron^{-1} \leq x \leq 8~\micron^{-1}$ to represent
the entire range from $3.3~\micron^{-1} \leq x \leq 11~\micron^{-1}$.
The term CCM89 used for $8~\micron^{-1} \leq x \leq 11~\micron^{-1}$
was an extrapolation of the FM90 far-UV term.
Thus, for the entire wavelength range considered in this paper
($3.3~\micron^{-1} \leq x \leq 11~\micron^{-1}$) the functional form
we use for our $R(V)$ dependent relationship is
\begin{eqnarray}
a(x) & = & 1.896 - 0.372x - 0.0108/[(x - 4.57)^2 + 0.0422] \nonumber \\
 & & + F_a(x) \\
b(x) & = & -3.503 + 2.057x + 0.718/[(x - 4.59)^2 + 0.0530] \nonumber \\
 & & + F_b(x)
\end{eqnarray}
where for ($5.9 \leq x \leq 8$)
\begin{eqnarray}
F_a(x) & = & -0.110(x - 5.9)^2  - 0.0099(x - 5.9)^3  \\
F_b(x) & = & 0.537(x - 5.9)^2  + 0.0530(x - 5.9)^3 
\end{eqnarray}
and for ($x < 5.9$)
\begin{equation}
F_a(x) = F_b(x) = 0.
\end{equation}

\begin{figure*}[tbp]
\epsscale{0.9}
\plotone{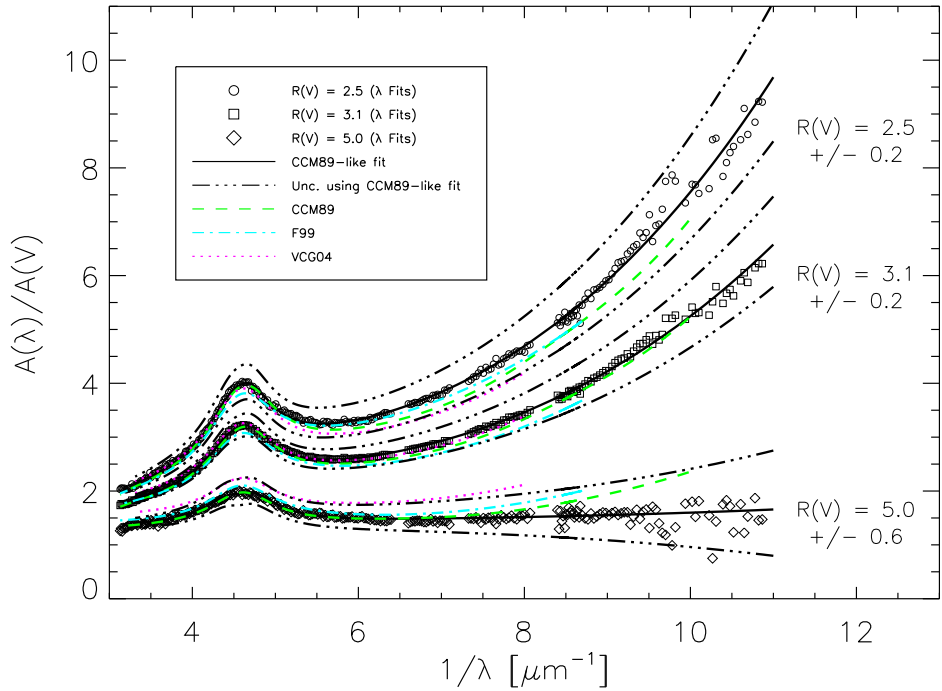}
\caption{The average extinction curves for $R(V)$ values of 2.5, 3.1,
and 5.0 are shown.  The new average curves derived in this paper are
shown as well as those from CCM89, F99, and VCG04.  The uncertainty on
our $R(V)$ dependent relationship (i.e., $a(x)$ and $b(x)$) is shown
using the $R(V)$ dependent fit that best describes the behavior of the
uncertainties on the fits at each individual wavelength.  In general,
the fit uncertainties translate to an equivalent uncertainty of
approximately 10\% on $R(V)$.
\label{fig_fuse_ccm_old_new}}
\end{figure*}

We compare the average curves calculated from our newly derived $R(V)$
dependent relationship with those given by CCM89, F99, and VCG04 in
Fig.~\ref{fig_fuse_ccm_old_new}.  The $R(V)$ dependent relationship we
derive is not noiseless and we show the uncertainties on the average
curves to illustrate this point.  The uncertainties on the curves
resulting from uncertainties on $a(x)$ and $b(x)$ can be approximated
by uncertainties on the value of $R(V)$.  Overall, these uncertainties
are equivalent to approximately a 10\% error on $R(V)$.  The
uncertainties are significant and grow with decreasing wavelength.
Overall, our new $R(V)$ dependent relationship is roughly consistent
within our uncertainties with past measurements in areas of common
wavelength coverage.  But, over most of the ultraviolet we do find a
stronger dependence on $R(V)$ than previous works.  The statistically
significance of this stronger dependence is difficult to assess given
that previous works did not include an analysis of the uncertainties
in their derivations of the $a(x)$ and $b(x)$.  For VCG04, some of the
differences can be traced to this study using unweighted linear least
square fits to determine $a(x)$ and $b(x)$.  We checked this effect
and found much shallower $R(V)$ dependences with our sample if we
perform unweighted fits.  Taking into account the uncertainties on
both $A(\lambda)/A(V)$ and $R(V)$ is clearly important.  For CCM89 and
F99, the largest differences are seen for $x > 7~\micron^{-1}$ where
we find a stronger dependence on $R(V)$.  This 
is not surprising as only with this new work has this wavelength range
been systematically studied.

\subsection{Structure}
\label{sec_structure}

The existence of structure in the UV extinction curve can give strong
clues to the carriers of different dust grain components.  The main
UV absorption
feature is the broad 2175~\AA\ bump which has been identified with
carbonaceous grains \citep{Stecher65graphite, Draine93}.  The other
main structure is the far-UV rise 
and this feature is also identified (but with less confidence) with
carbonaceous grains \citep{Joblin92, Li01}.  Other than these two features,
no other convincing UV features have been found.
\citet{Clayton03UVDIBs} has presented the most sensitive search for UV
absorptions to date.  They observed two heavily reddened sightlines
and derived a 
3$\sigma$ upper limit of $\sim$$0.02 A(V)$ on any features 20~\AA\ or
wider.  This result has been strengthened by \citet{Fitzpatrick07} who
found a 3$\sigma$ upper limit of $\sim$$0.06 A(V)$ on features 10~\AA\
or wider from the average residuals of 318 extinction curves.

\begin{figure*}[tbp]
\epsscale{1.2}
\plotone{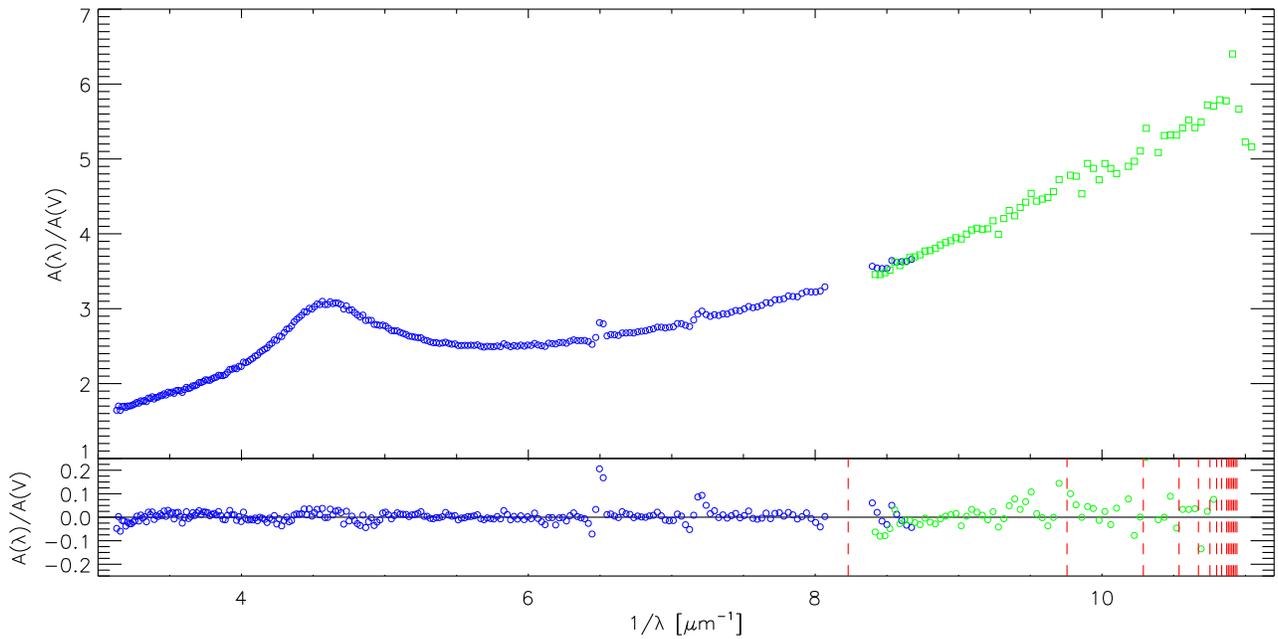}
\caption{The average extinction curve for the entire sample of 75
sightlines is plotted along with the average residual curve at a
resolution of 250.  The average residual curve was created by
averaging the 75 individual observed minus FM90 fit curves.  The
deviations seen at 6.5 and 7.5~\micron$^{-1}$ are the result of ISM or
wind line mismatches.
\label{fig_fuse_ave}}
\end{figure*}

\begin{figure*}[tbp]
\epsscale{1.2}
\plotone{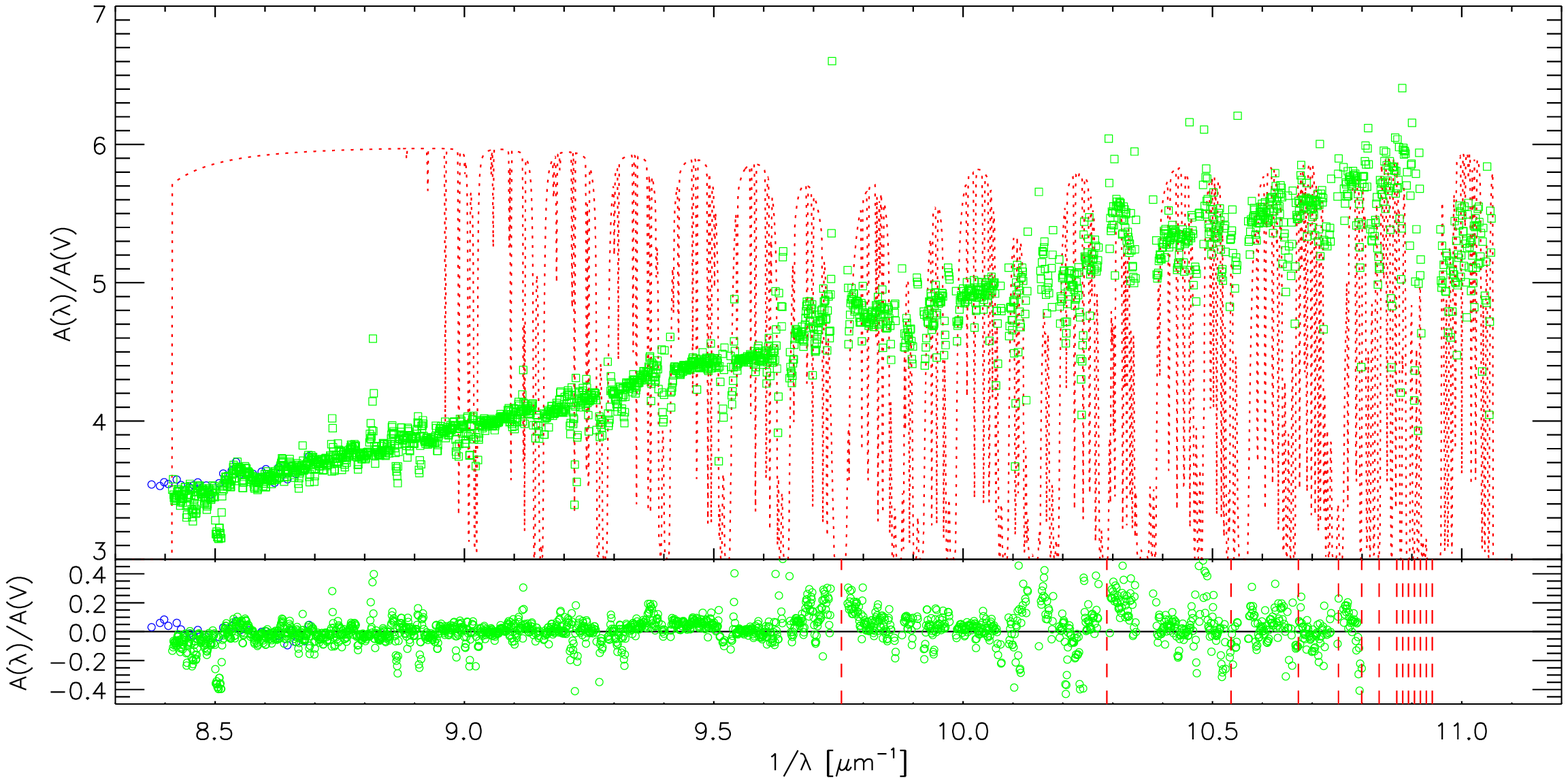}
\caption{The average extinction curve for the entire sample of 75
sightlines is plotted along with the average residual curve at a
resolution of $10^4$.  Overplotted on the average extinction curve is
the $H_2$ and \ion{H}{1} absorption spectrum (dotted line) for HD~122879 ($A(V)
= 1.41, R(V) = 3.17$) to provide a guide to where strong $H_2$ and \ion{H}{1}
absorptions occur.  The average residual curve was created by averaging 75
individual observed minus FM90 fit curves.  Overplotted on the residual
plot as dashed vertical lines is the location of the \ion{H}{1} absorption
features.
\label{fig_fuse_ave_fullres}}
\end{figure*}

We have used our sample of 75 FUSE extinction curves to search of
structure in the far-UV region.  Fig.~\ref{fig_fuse_ave} presents the
average extinction curve and residual curve for the entire sample at a
spectral resolution of 250.  There are two obvious features in the IUE
spectral region and these correspond to either ISM or wind line
mismatches between the reddened and comparison stars.  There are no
obvious features in the FUSE spectral region 
other than the scatter getting larger around $x = 9.25~\micron^{-1}$ and
then getting much larger after $x = 10.75~\micron^{-1}$.  The scatter of
the FUSE residuals puts a 3$\sigma$ upper limit of $\sim$$0.12 A(V)$
on features with a resolution of 250 ($\sim$4~\AA\ width).  For
comparison to previous studies, our 3$\sigma$ upper limit in the IUE
range is $\sim$$0.04 A(V)$ on features with a resolution of 250
($\sim$8~\AA width).

\begin{figure*}[tbp]
\epsscale{1.2}
\plotone{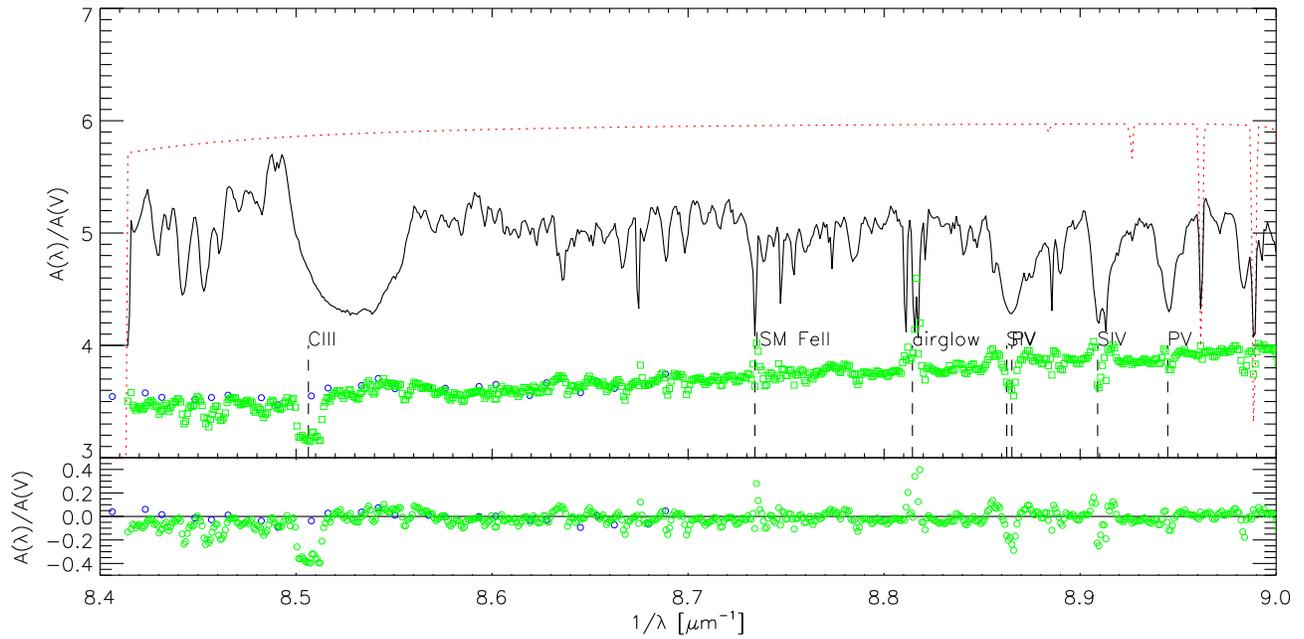}
\caption{The average extinction curve for $8.4 < \lambda^{-1} <
9.0~\micron^{-1}$ for the entire sample of 75
sightlines is plotted along with the average residual curve at a
resolution of $10^4$.  Overplotted on the average extinction curve is
the spectrum of HD~122879 and the $H_2$ and \ion{H}{1} absorption
spectrum for this sightline ($A(V) = 1.41, R(V) 
= 3.17$).  The average residual curve was created by averaging 75
individual observed minus FM90 fit curves.  The locations of features
that produce residuals in the average extinction curve are labeled.
\label{fig_fuse_ave_fullres_blowup}}
\end{figure*}

We can also search for features at significantly higher resolution
given that the native resolution of our FUSE spectra are on order
$10^4$.  Fig.~\ref{fig_fuse_ave_fullres} presents the average FUSE
extinction curve and residual curve for the entire sample at a
spectral resolution of $10^4$.  There are a number of features in the
residual curve.  These features can be seen to correlate with strong
absorptions in the example $H_2$ and \ion{H}{1} absorption spectrum.  For
example, two of the largest deviations correspond to Ly$\beta$ and
Ly$\gamma$ at 9.76 and 10.29~\micron$^{-1}$, respectively.  In
addition, there is a strong residual at $\sim$10.15~\micron$^{-1}$
that corresponds to a particularly strong $H_2$ absorption complex.
In general, all of the strong residuals in the region beyond
8.9~\micron$^{-1}$ correspond to regions of strong \ion{H}{1} or $H_2$
absorption.  There are residuals at $x < 9~\micron^{-1}$ that
are not associated with \ion{H}{1} or $H_2$ absorptions.
Fig.~\ref{fig_fuse_ave_fullres_blowup} shows a blowup of this
wavelength region including the spectrum of HD~122879 as a guide to
the origin of the stronger residuals.  The strong residuals all
correspond to strong stellar (\ion{C}{3}, \ion{S}{4}, and \ion{P}{5}),
ISM (\ion{Fe}{2}), or airglow lines \citep{Pellerin02}.  Examining
Fig.~\ref{fig_fuse_ave_fullres}, the scatter of 
the FUSE full resolution residuals clearly becomes large where
$x \sim 9.6~\micron^{-1}$.  Thus, the 3$\sigma$ upper limits are
$\sim$$0.15 A(V)$ for $x < 9.6$~\micron$^{-1}$ and $\sim$$0.68
A(V)$ for $x > 9.6$~\micron$^{-1}$ on features with a resolution
of $10^4$ ($\sim$0.1~\AA\ width).

\begin{figure*}[tbp]
\epsscale{1.2}
\plotone{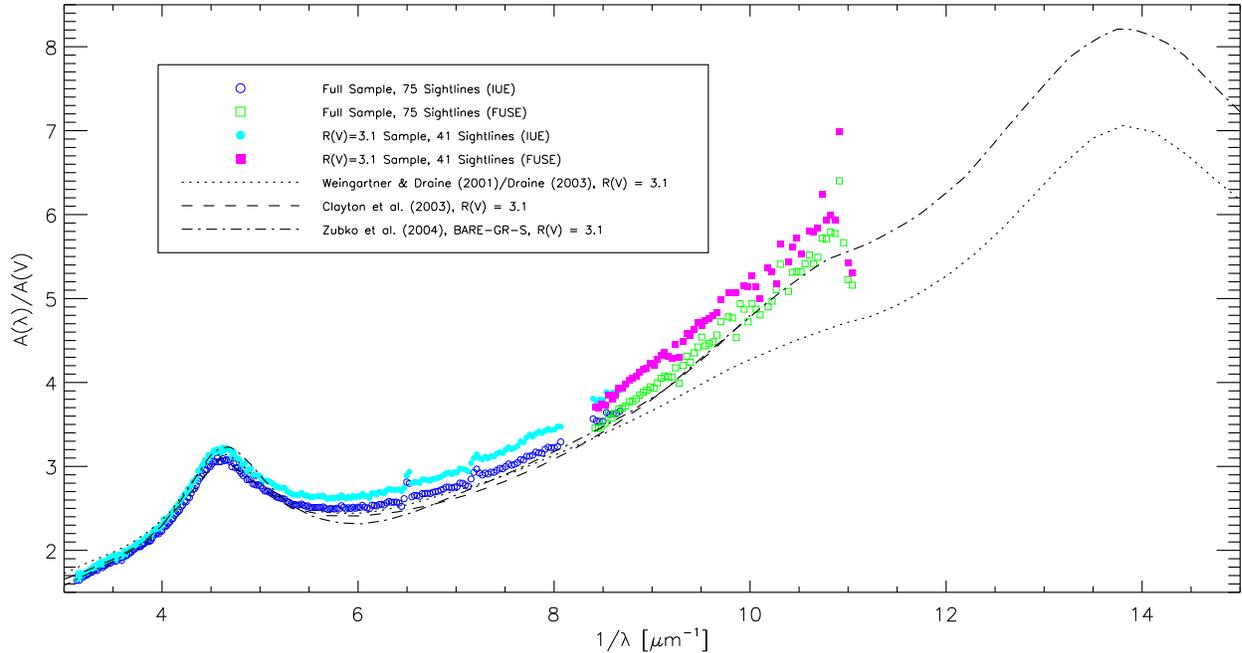}
\caption{The average extinction curve for the 41 curves in the sample
that have $R(V)$ values within $1\sigma$ of 3.1 is shown at a
resolution of 250.  The $R(V) = 3.1$ models from \citet{Weingartner01},
\citet{Clayton03}, and \citet{Zubko04} are plotted for comparison.
Note that the \citet{Clayton03} model only extends to
$10~\micron^{-1}$.
\label{fig_fuse_ave31}}
\end{figure*}

The average far-UV extinction curve can be compared with existing dust
grain models to evaluate how well such grain models do at describing
and/or predicting the far-UV extinction.  The comparison is shown in
Fig.~\ref{fig_fuse_ave31} where we show the average of all 75
sightlines (same as in Figs.~\ref{fig_fuse_ave} \&
\ref{fig_fuse_ave_fullres}) and the average of the 41 sightlines that
have $R(V)$ values within $1\sigma$ of 3.1.  The average $R(V)$ for
the full sample is 3.33 while the average for the $R(V) = 3.1$ sample
of 41 sightlines is, not surprisingly, 3.13.  In addition to the
observed average curves, three
different dust grain models \citep{Weingartner01, Clayton03,
Zubko04} are shown in this figure.  These models used different
constraints and methods to 
determine the dust grain size distribution and have known differences
as a result.  While all of the dust grain models shown are for an
$R(V) = 3.1$ (diffuse ISM), they are in closer agreement with the
average for our full sample than the $R(V) = 3.1$ average.  This
difference can be mainly traced to the differences seen between our
work and previous determinations of the $R(V)$ dependent relationship.
\citet{Weingartner01} and \citet{Zubko04} used the $R(V)=3.1$ curve from
F99 while \citet{Clayton03} used the $R(V)=3.1$ curve
from CCM89.  As both the CCM89 and F99 $R(V)=3.1$ curves are below our
$R(V)=3.1$ determined from the $R(V)$ dependent relationship (see
Fig.~\ref{fig_fuse_ccm_old_new}), it is not surprising the dust grain
models are also below our curve.  The differences between the CCM89,
F99, and this work are due to different samples of extinction curves
used as well as different fitting approaches.  The \citet{Clayton03} and
\citet{Zubko04} models do 
a much better job in reproducing the far-UV extinction levels than the
\citet{Weingartner01} \citep[as updated by][]{Draine03review} model.
Overall, the far-UV rise seems to consistent with its potential origin as a
second resonance feature centered around 14~$\micron^{-1}$
\citep{Joblin92, Li01}.

\section{Conclusions}

We have presented FUSE+IUE extinction curve for 75 sightlines that
have $A(V)$ values from 0.6 to 3.2 mag.  The $R(V)$ values of this
sample range from 2.5 to 5.5 with a strong clustering around 3.  These
extinction curves were created using the standard pair method.  Given
the FUSE sensitivities, this required generating a new set of
unreddened comparison stars.  This set of 75 FUSE extinction curves
represents a large increase in the number of extinction curves well
measured in the far-ultraviolet.

Using this sample, we investigated the nature of the far-UV
extinction with three methods. The first was to fit all the curves
with the FM90 parameterization.  We found that the FM90 parameters
determined with just the IUE portion of the extinction curve to be
broadly consistent with those that were determined using the FUSE+IUE
extinction curves.  There were significant changes in the individual
values of the $C_4^{A(V)}$
and $\gamma$ parameters.  In addition, the average $C_4^{A(V)}$
parameter decreased by $\sim$8\% indicating that the strength of the
far-UV rise is overestimated when using only IUE data.  All
the extinction curves had positive far-UV rises when fit with the
FUSE+IUE data giving no indication that the far-UV rise was turning
over at the shortest wavelengths.  We tested 
the new FM07 parameterization (with 1 additional parameter and a
simplified far-UV rise term) using the combined FUSE+IUE curves and
found only a weak preference for the FM07 parameterization.  We
choose to use the FM90 parameterization as it is simpler and provides
for direct comparisons to previous work.

The second investigation of the far-UV extinction concentrated on
deriving the $R(V)$ dependent relationship.  We derived the linear
dependence 
of $A(\lambda)/A(V)$ on $R(V)^{-1}$ at each FUSE and IUE wavelength
including the
observational uncertainties as part of the fitting.  We found a
stronger dependence of $A(\lambda)/A(V)$ on $R(V)^{-1}$ than
previously indicated in the far-UV.  It is not surprising as the
previous work in the far-UV was based on only a
very limited number of sightlines.  We include the uncertainties in
the fit in the $R(V)$ dependent parameterization and find that our
derivation is broadly consistent with previous work \citep{Cardelli89,
Fitzpatrick99, Valencic04} in the regions of overlap (mainly the IUE
range).  We present our $R(V)$ dependent relationship for the entire
UV ($3.3~\micron^{-1} \le \lambda^{-1} \le 11~\micron^{-1}$) using the same
functional form as used by CCM89.

Finally, we searched for discrete absorption features in the far-UV.
We looked for features at a resolution of 250 by averaging all 75
extinction curves and also averaging the residuals of each curve from
its corresponding FM90 fit.  We find a 3$\sigma$ upper limit of
$\sim$$0.12 A(V)$ on features with a resolution of 250 ($\sim$4~\AA\
width).  Utilizing the high spectral resolution nature of the FUSE
observations, we also searched for features at the native resolution
of $10^4$.  We found 3$\sigma$ upper limits of $\sim$$0.15 A(V)$ for
$\lambda^{-1} < 9.6$~\micron$^{-1}$ and $\sim$$0.68 A(V)$ for $\lambda^{-1} >
9.6$~\micron$^{-1}$ on features with a resolution of $10^4$
($\sim$0.1~\AA\ width).

\acknowledgements
We thank Derck Massa for useful conversations on extinction curve
uncertainties and Karl Misselt for discussion of dust grain models.
We appreciated the comments of the anonymous referee that resulted in
a stronger paper.  This study was supported by NASA ADP grant
NAG5-13033.  Based on observations made with the NASA-CNES-CSA Far
Ultraviolet Spectroscopic Explorer. FUSE is operated for NASA by the
Johns Hopkins University under NASA contract NAS5-32985.

\end{document}